\documentclass[12pt]{article}
\usepackage{graphicx}
\usepackage{rotating}
\renewcommand{\baselinestretch}{1.}
\begin{document}
\begin{center}
{\Large\bf Kinetic freeze-out temperature and flow velocity
extracted from transverse momentum spectra of final-state light
flavor particles produced in collisions at RHIC and LHC}

\vskip0.75cm

Hua-Rong Wei$^{a}$, Fu-Hu Liu$^{a,}${\footnote{E-mail:
fuhuliu@163.com; fuhuliu@sxu.edu.cn}}, and Roy A.
Lacey$^{b,}${\footnote{E-mail: Roy.Lacey@Stonybrook.edu}}

\vskip0.25cm

{\small\it $^a$Institute of Theoretical Physics, Shanxi
University, Taiyuan, Shanxi 030006, China

$^b$Departments of Chemistry \& Physics, Stony Brook University,
Stony Brook, NY 11794, USA}
\end{center}

\vskip0.5cm

{\bf Abstract:} The transverse momentum spectra of final-state
light flavor particles produced in proton-proton ($p$-$p$),
copper-copper (Cu-Cu), gold-gold (Au-Au), lead-lead (Pb-Pb), and
proton-lead ($p$-Pb) collisions for different centralities at
relativistic heavy ion collider (RHIC) and large hadron collider
(LHC) energies are studied in the framework of a multisource
thermal model. The experimental data measured by the STAR, CMS,
and ALICE Collaborations are consistent with the results
calculated by the multi-component Erlang distribution and Tsallis
Statistics. The effective temperature and real temperature
(kinetic freeze-out temperature) of interacting system at the
stage of kinetic freeze-out, the mean transverse flow velocity and
mean flow velocity of particles, and the relationships between
them are extracted. The dependences of effective temperature and
mean (transverse) momentum on rest mass, moving mass, centrality,
and center-of-mass energy, and the dependences of kinetic
freeze-out temperature and mean (transverse) flow velocity on
centrality, center-of-mass energy, and system size are obtained.
\\

\\


\vskip1.0cm

{\section{Introduction}}

Since the relativistic heavy ion collider (RHIC) and large hadron
collider (LHC) successfully run, the evolution process of
collision system and properties of quark-gluon plasma (QGP) [1]
formed in a high-temperature and high-density extreme condition,
attract more interests and studies. Many theoretical and
experimental methods are used to study the high energy
nucleus-nucleus collisions which result in multi-particle
productions. The model analysis is a simple but effective method
in the study of high energy nucleus-nucleus collisions. It can
extract the information of interacting system and QGP by analyzing
various spectra of final-state products with distribution laws of
different models. For example, by using the blast-wave model [2],
thermal and statistical model [3], Landau hydrodynamic model
[4--7], multisource thermal model [8--10], and so forth, one can
describe transverse momentum (mass) spectrum, azimuthal
distribution, and rapidity distribution of final-state products to
extract temperature, non-equilibrium degree, longitudinal
extension, and speed of sound of interacting system, and flow
velocity and chemical potential of particles [11--14], especially
at the state of kinetic freeze-out, which renders that model
analyzes have made great contributions to study the properties of
reaction system and QGP, as well as new physics.

In high energy collisions, the interacting system at the kinetic
freeze-out (the last stage of collisions) stays at a thermodynamic
equilibrium state or local equilibrium state, when the particle
emission process is influenced by not only the thermal motion but
also the flow effect. In other words, it is interesting to study
the temperature of interacting system and flow velocity of
particles at the stage of kinetic freeze-out. The effective
temperature extracted from the transverse momentum spectrum
[15--25] includes thermal motion and flow effect of particles,
where the thermal motion is actually the reflection of the real
temperature of emission source. From dissecting the effective
temperature, it is possible to obtain the real temperature of
interacting system (kinetic freeze-out temperature) and mean
(transverse) flow velocity of particles. The relationships between
effective temperature, real temperature, flow velocity are not
totally clear. Although the theories of studying kinetic
freeze-out temperature are many, their results are different from
each other in some cases. This indicates that studying more their
relations is important and needful.

In the present work, we use the Erlang distribution with one-,
two-, or three-component and Tsallis statistics in the multisource
thermal model [8--10] to describe the transverse momentum spectra
of final-state particles produced in proton-proton ($p$-$p$),
copper-copper (Cu-Cu), gold-gold (Au-Au), lead-lead (Pb-Pb), and
proton-lead ($p$-Pb) collisions with different centrality
intervals over a $\sqrt{s_{NN}}$ (center-of-mass energy per
nucleon pair) or $\sqrt{s}$ (only for $p$-$p$ collision in some
cases) range from 0.2 to 7 TeV. The Monte Carlo method is used to
calculate the results and to see the statistical fluctuations in
the process of calculation. The calculated results are compared
with the experimental data of the STAR [15--19], CMS [20], and
ALICE [21--25] Collaborations. From comparison and analysis, the
kinetic freeze-out temperature of interacting system, mean
transverse flow velocity and mean flow velocity of particles, and
their relations are then extracted.

The structure of the present work is as followings. The model and
formulism are shortly described in section 2. Results and
discussion are given in section 3. In section 4, we summarize our
main observations and conclusions.
\\

{\section{The model and formulism}}

The present work is based on the multisource thermal model
[8--10], which assumes that many emission sources are formed in
high energy collisions. Due to the existent of different
interacting mechanisms in collisions and measurement of different
event samples in experiments, these sources are classified into a
few groups. Generally, we assume that sources in the same group
stay at a local equilibrium state, which means they have the same
excitation degree and a common temperature. The emission process
of all the sources in different groups results in the final-state
spectrum, which can be described by a multi-component distribution
law, because of a local equilibrium state corresponding to a
singular distribution. Using different distributions to describe
transverse momentum ($p_T$) spectra of final-state particles can
obtain different information. For example, a multi-component
Erlang $p_T$ distribution can directly give the mean transverse
momentum of each group, while the Tsallis statistics can show the
effective temperature of the whole interacting system which may
have group-by-group fluctuations in different local thermal
equilibriums.

According to the model [10], particles generated from one emission
source is assumed to obey an exponential function of transverse
momentum distribution
\begin{equation}
f_{ij}(p_{tij})=\frac{1}{\langle{p_{tij}}\rangle}
\exp{\bigg[-\frac{p_{tij}}{\langle{p_{tij}}\rangle}\bigg]},
\end{equation}
where $p_{tij}$ is the transverse momentum contributed by the
$i$th source in the $j$th group, and $\langle{p_{tij}}\rangle$ is
the mean value of $p_{tij}$. All the sources in the $j$th group
results in the folding of exponential functions
\begin{equation}
f_{j}(p_{T})=\frac{p_T^{m_{j}-1}}{(m_{j}-1)!
\langle{p_{tij}}\rangle^{m_{j}}} \exp{\bigg[
-\frac{p_{T}}{\langle{p_{tij}}\rangle}\bigg]},
\end{equation}
where $m_{j}$ is the source number in the $j$th group, and $p_{T}$
denotes the transverse momentum contributed by the $m_{j}$
sources. This is the Erlang distribution. The contribution of all
$l$ groups of sources can be expressed as
\begin{equation}
f_E(p_{T})=\sum_{j=1}^{l}k_{j}f_{j}(p_{T}),
\end{equation}
where $k_{j}$ denotes the relative weight contributed by the $j$th
group and meets the normalization $\sum_{j=1}^{l}k_{j}=1$. This is
the multi-component Erlang distribution. Then, using the inverse
slope parameter $\langle{p_{tij}}\rangle$, we can obtain the mean
transverse momentum $\langle p_{T}\rangle$ of final-state
particles to be
\begin{equation}
\langle p_{T}\rangle=\sum_{j=1}^{l}k_{j}m_{j}\langle{p_{tij}}\rangle.
\end{equation}

The Tsallis statistics is in fact the sum of contributions of two
or three standard distributions. Although the Tsallis statistics
does not give each local temperature like multi-component
distribution which reveals fluctuations from a local equilibrium
state to another one, it use an average temperature of the whole
interacting system to describe the effect of local temperature
fluctuations. So, the Tsallis statistics is widely used in high
energy collisions [26--35].

According to the Tsallis statistics [26--31], we use the formalism
of unit-density function of $p_T$ and rapidity ($y$)
\begin{equation}
\frac{d^2N}{dydp_T}=Cp_T\sqrt{p_T^2+m_0^2}\cosh y
\bigg[1+\frac{q-1}{T} \sqrt{p_T^2+m_0^2} \cosh y
\bigg]^{-\frac{q}{q-1}},
\end{equation}
where $N$ is the number of particles, $C= gV/(2\pi)^2$ is the
normalization constant, $g$ and $V$ are degeneracy factor and
volume respectively, $m_0$ is the rest mass of considered
particle, $T$ is the mean effective temperature over fluctuations
in different groups, and $q$ ($q>1$) is the factor (entropy index)
to characterize the degree of non-equilibrium among different
groups. With an integral for $y$ in Eq. (5), the normalized
Tsallis $p_T$ distribution is obtained and can be written as
\begin{equation}
f_{T}(p_T)=\frac{1}{N}\frac{dN}{dp_T}=C_{T}
p_T\sqrt{p_T^2+m_0^2}\int_{y_{\min}}^{y_{\max}} \cosh
y\bigg[1+\frac{q-1}{T}\sqrt{p_T^2+m_0^2} \cosh
y\bigg]^{-\frac{q}{q-1}} dy,
\end{equation}
where $C_{T}$ denotes the normalization constant which results in
$\int_0^{\infty} f_{T}(p_T) dp_T=1$, $y_{\min}$ is the minimum
rapidity, and $y_{\max}$ is the maximum rapidity.

Based on the above two $p_T$ distributions, we can use the Monte
Carlo method to obtain a series of $p_T$. Under the assumption of
isotropic emission in the source rest frame, the space angle
$\theta$ and azimuthal angel $\phi$ of particles satisfy the
distributions of $f_{\theta}(\theta)=(1/2) \sin \theta$ and
$f_{\phi}(\phi)=1/(2\pi)$ respectively. By the Monte Carlo method,
a series of $\theta$ and $\phi$ are obtained. Correspondingly, the
$x$-component, $y$-component, and (longitudinal) $z$-component of
momentum are $p_x=p_T\cos\phi$, $p_y=p_T\sin\phi$, and
$p_z=p_T/\tan \theta$, respectively. Then, the momentum
$p=p_T/\sin \theta$ or $p=\sqrt{p_T^2+p^2_z}$, the energy
$E=\sqrt{p^2+m_0^2}$, the Lorentz factor
$\gamma=1/\sqrt{1-(p/E)^2}$, and the moving mass $m=m_{0}\gamma$,
as well as their averages $\langle{p_{T}}\rangle$,
$\langle{p}\rangle$, $\overline{\gamma}$, and $\overline{m}$ are
acquired. Particularly, the values of mean transverse momentums
$\langle{p_{T}}\rangle$ according to analytical function and Monte
Carlo method are almost the same.
\\

{\section{Results and discussion}}

Figure 1 presents the transverse momentum spectra of various
identified hadrons produced in $p$-$p$ collision at center-of-mass
energy (a) $\sqrt{s}=0.2$, (b) (c) $0.9$, (d) $2.76$, and (e) (f)
$7$ TeV, where $N_{EV}$ on the axis denotes the number of
inelastic collisions events. The symbols represent the
experimental data of (a) $\pi^{+}$, $K^{+}$, and $p$ measured by
the STAR Collaboration at midrapidity $|y|<0.1$ [15], (b) (d) (e)
$\pi^{+}$, $K^{+}$, and $p$ measured by the CMS Collaboration in
the range $|y|<1$ [20], (c) $\Lambda$, $\phi$, and
$\Xi^{-}+\bar{\Xi}^{+}$ measured by the ALICE Collaboration in the
range $|y|<0.8$ [21], as well as (f) $\Xi$ and $\Omega$ measured
by the ALICE Collaboration in the range $|y|<0.5$ [22]. The errors
include both the statistical and systematic errors. The solid and
dashed curves are our results calculated by using the one- or
two-component Erlang distribution and Tsallis statistics,
respectively. The values of free parameters ($m_{1}$, $p_{ti1}$,
$k_{1}$, $m_{2}$, and $p_{ti2}$), normalization constant ($N_{\rm
E0}$), and $\chi^2$ per degree of freedom ($\chi^2$/dof)
corresponding to the one- or two-component Erlang distribution are
listed in Table 1, and the values of free parameters ($T$ and
$q$), normalization constant ($N_{\rm T0}$), and $\chi^2$/dof
corresponding to the Tsallis statistics are given in Table 2,
where the normalization constants ($N_{\rm E0}$ and $N_{\rm T0}$)
are used for comparisons between the normalized curves and data
points. In particular, the value of $\chi^2$/dof for
$\Xi^{-}+\bar{\Xi}^{+}$ in Figure 1(c) is in fact the value of
$\chi^2$ due to the number of data points being less than that of
parameters. One can see that the one- or two-component Erlang
distribution and Tsallis statistics describe the experimental data
of the considered particles in $p$-$p$ collision at different
energies. From Tables 1 and 2, one can see that the numbers of
sources in different groups are 2, 3, or 4. The effective
temperature $T$ increases and the non-equilibrium degree parameter
$q$ decreases with increase of rest mass, which reflects
non-simultaneous productions of different types of particles,
while $T$ and $q$ increase with increase of center-of-mass energy.
The normalization constants $N_{\rm E0}$ and $N_{\rm T0}$ decrease
with increase of rest mass, and increase with increase of
center-of-mass energy. It is interesting to note that the product
of $T$ and $q$ increases with increase of rest mass and
center-of-mass energy.

Figure 2 shows the transverse momentum spectra of (a) $K_{S}^{0}$,
(b) $\Lambda$, (c) $\Xi$, and (d) $\Omega$ produced in Cu-Cu
collisions at $\sqrt{s_{NN}}=0.2$ TeV. The symbols represent the
experimental data of the STAR Collaboration in $|y|<0.5$ and
different centrality ($C$) intervals of 0--10\%, 20--30\%, and
40--60\% [16]. The error bars are combined statistical and
systematic errors. The solid and dashed curves are our results
calculated by using the one- or two-component Erlang distribution
and Tsallis statistics, respectively. For clarity, the results for
different intervals are scaled by different amounts shown in the
panels. The values of free parameters, normalization constants,
and $\chi^2$/dof are displayed in Tables 1 and 2. Obviously, the
one- or two-component Erlang distribution and Tsallis statistics
describe well the experimental data of the considered particles in
0.2 TeV Cu-Cu collisions with different centrality intervals. The
numbers of sources in different groups are 1, 2, 3, or 4. The
parameter $T$ increases and the parameter $q$ decreases with
increases of rest mass and centrality. The product of $T$ and $q$
increases with increases of rest mass and centrality. The
parameters $N_{\rm E0}$ and $N_{\rm T0}$ decrease with increases
of rest mass and decrease of centrality.

The $p_{T}$ spectra of (a) $\pi^+$ for $|y|<0.5$, (b) $K_{S}^{0}$
for $|y|<0.5$, (c) $p$ for $|y|<0.5$, (d) $\phi$ for $|y|<0.5$,
(e) $\Lambda$ for $|y|<0.5$, and (f) $\Xi^{-}$ for $|y|<0.75$
produced in Au-Au collisions at $\sqrt{s_{NN}}=0.2$ TeV as a
function of centrality are given in Figure 3. The experimental
data were recorded by the STAR Collaboration [17--19], and scale
factors for different centralities are applied to the spectra in
the panels for clarity. The uncertainties on the data points for
$\pi^+$, $K_{S}^{0}$, $p$, and $\Lambda$ are statistical and
systematic combined. While the uncertainties for $\phi$ and
$\Xi^{-}$ are only statistical uncertainties and systematic
uncertainties respectively. The results calculated by using the
one-, two-, or three-component Erlang distribution and Tsallis
statistics are shown in the solid and dashed curves, respectively.
The values of free parameters ($m_{1}$, $p_{ti1}$, $k_{1}$,
$m_{2}$, and $p_{ti2}$), normalization constant, and $\chi^2$/dof
corresponding to the one- or two-component Erlang distribution in
Figures 3(b)--3(f) are listed in Table 1. The values of free
parameters ($m_{1}$, $p_{ti1}$, $k_{1}$, $m_{2}$, $p_{ti2}$,
$k_{2}$, $m_{3}$, and $p_{ti3}$), normalization constant, and
$\chi^2$/dof corresponding to the three-component Erlang
distribution in Figure 3(a) are listed in Table 3. The values of
free parameters, normalization constant, and $\chi^2$/dof
corresponding to the Tsallis statistics are given in Table 2. Once
more, the two types of distributions are in good agreement with
the experimental data of the considered particles in 0.2 TeV Au-Au
collisions with different centrality intervals. The numbers of
sources in different groups are 1, 2, 3, or 4. The parameter $T$
increases and the parameter $q$ decreases with increases of rest
mass and centrality. The product of $T$ and $q$ increases with
increases of rest mass and centrality. The normalization constants
$N_{\rm E0}$ and $N_{\rm T0}$ decrease with increase of rest mass
and decrease of centrality.

The $p_{T}$ spectra of (a) $\pi^{+}$, (b) $K^{+}$, (c) $p$, and
(d) $\phi$ produced in central (0--5\%), semi-central (50--60\%),
and peripheral (80--90\%) Pb-Pb collisions at $\sqrt{s_{NN}}=2.76$
TeV are displayed in Figure 4. The symbols represent the
experimental data measured by the ALICE Collaboration at
midrapidity $|y|<0.5$ [23, 24]. The uncertainties on the data
points are combined statistical and systematic ones. The fitted
results with the one- or two-component Erlang distribution and
Tsallis statistics are plotted by the solid and dashed curves,
respectively. The values of free parameters, normalization
constants, and $\chi^2$/dof are exhibited in Tables 1 and 2. As
can be seen, the two types of distribution laws are consistent
with the experimental data of the considered particles in 2.76 TeV
Pb-Pb collisions with different centrality classes. The numbers of
sources in different groups are 1, 2, or 3. The effective
temperature $T$ increases with increases of rest mass and
centrality. The parameter $q$ decreases with increases of rest
mass and centrality. The product of $T$ and $q$ increases with
increases of rest mass and centrality. The parameters $N_{\rm E0}$
and $N_{\rm T0}$ decrease with increase of rest mass and decrease
of centrality.

Figure 5 exhibits the $p_{T}$ spectra of (a) $\pi^{+} + \pi^{-}$,
(b) $K^{+} + K^{-}$, (c) $p+\bar{p}$, and (d) $\Lambda
+\bar{\Lambda}$ produced in central (0--5\%), semi-central
(40--60\%), and peripheral (80--100\%) $p$-Pb collisions at
$\sqrt{s_{NN}}=5.02$ TeV. The ALICE experimental data in $0<y<0.5$
are represented by different symbols [25]. The error bars are
combined statistical and systematic errors. Our results analyzed
by the two-component Erlang distribution and Tsallis statistics
are given by the solid and dashed curves, respectively. The values
of free parameters, normalization constants, and $\chi^2$/dof are
summarized in Tables 1 and 2. Once again, the experimental data
can be well described by the two types of fit functions for
$p_{T}$ spectra in all centrality bins. The numbers of sources in
different groups are 2, 3, or 4. The effective temperature $T$
increases with increases of rest mass and centrality. With
increases of rest mass and centrality, the parameters $q$, $N_{\rm
E0}$, and $N_{\rm T0}$ decrease, and the product of $T$ and $q$
increases.

In the above descriptions, it is easy to notice that the numbers
of sources $m_{j}$ in the $j$th group are in the range $1 \leq
m_{j} \leq 4$. The $m_{j}$ is so small that we think these sources
corresponding to a few partons which include sea and valent quarks
and gluons in high-energy collisions. Generally, the transverse
momentum spectrum is contributed by the sum of soft and hard
parts. The soft excitation process is a strong interacting process
where a few sea quarks and gluons taken part in, and the hard
scattering process is a more violent collision among a few valent
quarks in incident nucleons. In $p_{T}$ spectrum, the soft
excitation and hard scattering processes correspond to a narrow
low-$p_{T}$ and a wide high-$p_{T}$ regions respectively [36--38].
And in the describing by two-component Erlang distribution, the
first and second components correspond to the soft and hard
processes respectively. Although the low-$p_{T}$ region
contributed by soft process is narrow, the contribution of soft
excitation is main, which can be seen from the relative weight
$k_{1}>50\%$. Particularly, when the region of $p_{T}$ spectrum is
narrow enough, the two-component Erlang distribution actually is
the (one-component) Erlang distribution for the contribution of
the second component being neglected. On the contrary, when the
high-$p_{T}$ region is very wide, the hard process can not be
described by one component distribution, which means the
two-component distribution would expand to the three-component
Erlang distribution.

The temperature parameter $T$ extracted from the Tsallis
distribution is actually the effective temperature of emission
sources, which can not reflect the transverse excitation of
interacting system at the stage of kinetic freeze-out. In order to
obtain the real temperature (kinetic freeze-out temperature) of
emission sources and (transverse) flow velocity of final-state
particles, we study the linear dependences of mean transverse
momentum $\langle p_{T}\rangle$, mean momentum $\langle p
\rangle$, and effective temperatures $T$ on particle rest mass
$m_{0}$ and mean moving mass $\overline{m}$. Figures 6--8 show the
center-of-mass energy and centrality dependences of $\langle
p_{T}\rangle$, $\langle p \rangle$, and $T$ on $m_{0}$
respectively, and Figures 9--11 show that on $\overline{m}$
respectively. One can see that the three quantities $\langle
p_{T}\rangle$, $\langle p \rangle$, and $T$ increase with increase
of $m_{0}$ and $\overline{m}$. From $p$-$p$ collision, the three
quantities increase with increase of center-of-mass energy, and
from Cu-Cu, Au-Au, Pb-Pb, and $p$-Pb collisions, the three
quantities increase with increase of centrality.

To see clearly the dependences of $\langle p_{T}\rangle$, $\langle
p \rangle$, and $T$ on $m_{0}$ and $\overline{m}$, we fit the
linear relationships which can be written as
\begin{equation}
\langle p_{T} \rangle=C_{0}+k_{T}m_{0},
\end{equation}
\begin{equation}
\langle p \rangle=C_{p}+k_{p}m_{0},
\end{equation}
\begin{equation}
T=T_{0}+km_{0},
\end{equation}
\begin{equation}
\langle p_{T} \rangle=C_{0}^{'}+\langle u_{T} \rangle
\overline{m},
\end{equation}
\begin{equation}
\langle p \rangle=C_{p}^{'}+\langle u \rangle \overline{m},
\end{equation}
and
\begin{equation}
 T=C_{T}^{'}+k^{'} \overline{m},
\end{equation}
respectively, where the units of temperature, momentum, velocity,
and mass are GeV, GeV/$c$, $c$, and GeV/$c^2$, respectively, where
$c=1$ is in natural units; The intercepts ($C_{0}$, $C_{p}$,
$T_{0}$, $C_{0}^{'}$, $C_{p}^{'}$£¬and $C_{T}^{'}$) have the same
units as corresponding dependent variables; The slopes $k_{T}$,
$k_{p}$, $\langle u_{T} \rangle$, and $\langle u \rangle$ are in
the units of $c$, while $k$ and $k^{'}$ are in the units of $c^2$.
The values of intercepts, slopes, and $\chi^2$ are given in Table
4.

From Figures 6--11 and Table 4, we can see the intercepts and
slopes in different linear correlations. In all cases, the
intercepts with different center-of-mass energies ($p$-$p$
collision) or centrality bins (Cu-Cu, Au-Au, Pb-Pb, or $p$-Pb
collisions) have the tendency of converging to one point, which
means the intercepts are nearly equal to each other or do not
change obviously with center-of-mass energy or centrality, while
the slopes have the tendencies of increasing with center-of-mass
energy and centrality. The intercepts and slopes obtained from
$\langle p_{T}\rangle-m_{0}$, $\langle p \rangle-m_{0}$, and
$T-m_{0}$ correlations are larger than those from $\langle
p_{T}\rangle-\overline{m}$, $\langle p \rangle-\overline{m}$, and
$T-\overline{m}$ correlations, but the changes about intercepts
are much larger than those about slopes. Besides, the increments
of slopes with center-of-mass energy or centrality in $\langle
p_{T}\rangle-m_{0}$, $\langle p \rangle-m_{0}$, and $T-m_{0}$
correlations are larger than those in $\langle
p_{T}\rangle-\overline{m}$, $\langle p \rangle-\overline{m}$, and
$T-\overline{m}$ correlations, which renders slow changes in the
latter cases.

In function $T=T_{0}+km_{0}$, the quantity $T$ extracted directly
from the distribution mentioned above, is the effective
temperature which includes thermal motion and flow effect of
particles. As the temperature corresponding to massless ($m_0=0$)
particle, $T_{0}$ is regarded as the source real temperature at
the kinetic freeze-out (or the kinetic freeze-out temperature of
interacting system) [39--43]. The flow effect of particles is
shown in quantity of $km_{0}$, where the slope $k$ has the
dimension of the square of velocity. At the same time, in function
$T=C_{T}^{'}+k^{'} \overline{m}$, although the intercept
$C_{T}^{'}$ has the same dimension as temperature, it's values
shown in Figure 11 and Table 4 approximately equal to zero, which
have no physics meaning. In correlations $\langle
p_{T}\rangle-m_{0}$, $\langle p_{T}\rangle-\overline{m}$, $\langle
p \rangle - m_{0}$, and $\langle p \rangle-\overline{m}$, the
slopes ($k_{T}$, $\langle u_{T} \rangle$, $k_{p}$, and $\langle u
\rangle$) have the dimension of velocity and are considered to be
related to mean transverse flow velocity and mean flow velocity.
It is interesting to find the relationships between the intercepts
and slopes, especially relationships between the several flow
velocities and $k$ (or $k^{'}$).

Figure 12 exhibits the correlations between intercepts (a)
$C_{p}-C_{0}$, (b) $C_{p}^{'}-C_{0}^{'}$, and (c) $T_{0}-C_{0}$ in
different collision systems with different centrality classes at
different energies, and corresponding fitting are executed. One
can see that $C_{p}$ and $C_{0}$, $C_{p}^{'}$ and $C_{0}^{'}$ have
explicit linear relations $C_{p}=(\pi/2)C_{0}$ and
$C_{p}^{'}=(\pi/2)C_{0}^{'}$ with almost zero $\chi^2$, which is
due to the assumption of isotropic emission in the source rest
frame. In $T_{0}-C_{0}$ correlation, the $T_{0}$ increases with
increase of $C_{0}$, and they are basically compatible with the
linear relation $T_{0}=(0.192\pm 0.073 )C_{0}$ with $\chi^2$/dof =
0.237. Other relations among intercepts do not show an obvious law
and are not shown in the panels due to trivialness.

At the same time, the correlations between different slopes (a)
$k_{p}-k_{T}$, (b) $k-k_{T}$, (c) $k_{T}-\langle u_{T}\rangle$,
(d) $\langle u \rangle-\langle u_{T}\rangle$, and (e)
$k^{'}-\langle u_{T}\rangle$ in different collision systems with
different centrality classes at different energies, as well as
corresponding fitting are shown in Figure 13. Once again, based on
the assumption of isotropic emission in the source rest frame, the
$k_{p}$ and $k_{T}$ from $\langle p \rangle-m_{0}$ and $\langle
p_{T} \rangle - m_{0}$, $\langle u \rangle$ and $\langle
u_{T}\rangle$ from $\langle p \rangle-\overline{m}$ and $\langle
p_{T} \rangle-\overline{m}$ also have the same explicit linear
relations $k_{p}=(\pi/2)k_{T}$ and $\langle u
\rangle=(\pi/2)\langle u_{T} \rangle$ with almost zero $\chi^2$.
From Figures 13(b) and 13(d), one can see that $k$ (or $k^{'}$)
increases with increase of $k_{T}$ (or $\langle u_{T} \rangle$),
and the parameters are in good agreement with the fitted power
function relations $k=[(1/3)\pm0.082]k_{T}^{2}$ with $\chi^2$/dof
= 1.546 and $k^{'}=[(1/2)\pm0.170]\langle u_{T} \rangle ^{2}$ with
$\chi^2$/dof = 1.472, respectively. The relationship between
$k_{T}$ and $\langle u_{T}\rangle$ respectively from $\langle
p_{T} \rangle - m_{0}$ and $\langle p_{T} \rangle - \overline{m}$
correlations is fitted by the line $k_{T}=(3.0\pm0.184)\langle
u_{T} \rangle -(0.714\pm0.082)$ with $\chi^2$/dof = 0.043. Other
relations among slopes do not show an obvious law and are not
shown in the panels due to trivialness. In addition, we would like
to point out that in the Monte Carlo calculation, some
conservation laws (such as energy conservation and momentum
conservation) and physics limitation (such as flow velocity $<c$)
are used so that we can obtain reasonable values and relations.

It is noticed that most of $\langle u_{T}\rangle$ values
(0.339--0.522$c$) extracted from $\langle p_{T}
\rangle-\overline{m}$ are slightly less than 0.5$c$ and $k_{T}$
(0.382--0.944$c$) from $\langle p_{T} \rangle-m_{0}$, while most
of $\langle u \rangle$ values (0.532--0.820$c$) extracted from
$\langle p \rangle-\overline{m}$ is more close to 0.5$c$ and less
than $k_{p}$ (0.546--1.424$c$) from $\langle p_{T} \rangle-m_{0}$.
Considering overlarge $k_T$ and $k_p$, we regard $\langle u_{T}
\rangle$ and $\langle u \rangle$ as the mean transverse flow
velocity and mean flow velocity of particles respectively, and
they obey the relation of $\langle u \rangle=(\pi/2)\langle u_{T}
\rangle$. In addition, $\langle u_{T} \rangle$ and $k^{'}$
approximatively obey $k^{'}=(1/2)\langle u_{T} \rangle ^{2}$.

The mean $T_{0}$ of interacting system obtained from $\langle T
\rangle-m_{0}$ does dot change in error range or approximates
independent of energy, centrality, particle type, and system size.
The mean $T_{0}$ corresponding to the Tsallis distribution is
($0.071\pm 0.007$) GeV, which is less than that [($0.137\pm
0.007$)] GeV from the Boltzmann distribution in previous work
[43], and is also less than that (0.177 GeV) from an exponential
shape of transverse mass spectrum [41]. Our result is close to
that (0.098 GeV) from the blast-wave model [44]. Comparing with
chemical freeze-out temperature, the kinetic freeze-out
temperature in present work is obviously less than that (0.170
GeV) of theoretical critical point of the QCD (quantum
chromodynamics) prediction [45--47], and that (0.156 GeV) from
particle ratios in a thermal and statistical model [3].

In the above discussions, we present in fact a method to extract
the kinetic freeze-out temperature and (transverse) flow velocity
from transverse momentum spectra in the multisource thermal model
[8--10], in which the sources are described by different
distribution laws. The method to extract the kinetic freeze-out
temperature is also used in other literature [39--43], and the
method to extract the (transverse) flow velocity is seldom
discussed elsewhere. Although the method used in the present work
is different from the blast-wave model [2] which discusses radial
flow, it offers anyhow another way to study the kinetic freeze-out
temperature and (transverse) flow velocity. In our recent work
[48], the multisource thermal model is revised with the blast-wave
picture, where we assume that the fragments and particles produced
by thermal reason in the sources are pushed away by a blast-wave.
The blast-wave causes the final-state products to be effected by
isotropic and anisotropic flows.
\\

{\section{Conclusions}}

From the above discussions, we obtain following conclusions.

(a) The transverse momentum distributions of final-state particles
produced in $p$-$p$, Cu-Cu, Au-Au, Pb-Pb, and $p$-Pb collisions
with different centrality bins over an energy range from 0.2 to 7
TeV, can be described by one-, two-, or three-component Erlang
distribution and Tsallis statistics in the framework of
multisource thermal model. The calculated results are consistent
with the experimental data of $\pi^{\pm}$, $K^{\pm}$, $K_S^0$,
$p$, $\bar p$, $\phi$, $\Lambda$, $\bar \Lambda$, $\Xi$, and
$\Omega$, etc. measured by the STAR, CMS, and ALICE
Collaborations.

(b) In most cases, the data of $p_T$ spectra are fitted by the
two-component Erlang distribution, where the first component
corresponding to a narrow low-$p_{T}$ region is contributed by the
soft excitation process in which a few sea quarks and gluons taken
part in and accounts for a larger proportion, and the second
component for a wide high-$p_T$ region indicates hard scattering
process which is a more violent collision among a few valent
quarks in incident nucleons. The mean transverse momentum $\langle
p_{T}\rangle$ extracted from multi-component Erlang distribution
increases with increases of center-of-mass energy, particle mass,
event centrality, and system size.

(c) The Tsallis $p_T$ distribution uses two free parameters $T$
and $q$ to describe the effective temperature and the
non-equilibrium degree of the interacting system respectively. The
present work shows that $q$ decreases with increases of particle
mass and event centrality, and increases with increase of
center-of-mass energy. In physics, $q=1$ corresponds to an
equilibrium state and a large $q$ corresponds to a state departing
far from equilibrium. Our study indicates that a high
center-of-mass energy results in interacting system deviating from
the equilibrium state, and the interacting system of central
collisions and heavier particles are closer to the equilibrium
state. The extracted $T$ increases with increases of
center-of-mass energy, particle mass, event centrality, and system
size.

(d) The intercept $T_{0}$ in $T-m_{0}$ correlation is regarded as
the mean kinetic freeze-out temperature of interacting system,
which do not depend on center-of-mass energy, event centrality,
and system size in error range, which renders that different
interacting systems at the stage of kinetic freeze-out stay at the
same phase state. The average of $T_{0}$ corresponding to the
Tsallis distribution is $(0.071\pm 0.007)$ GeV, which is less than
the effective temperature, chemical freeze-out temperature, and
expected critical temperature (130--165 MeV) of the QGP formation
[49].

(e) The mean transverse flow velocity $\langle u_{T} \rangle$ and
mean flow velocity $\langle u \rangle$ of particles are obtained
from $\langle p_{T} \rangle-\overline{m}$ and $\langle p
\rangle-\overline{m}$ correlations respectively. The present work
shows that $\langle u_{T} \rangle$ (0.339-0.522$c$) and $\langle u
\rangle$ (0.532-0.820$c$) have narrower range and are more close
to 0.5$c$ than those from $\langle p_{T} \rangle-m_{0}$ and
$\langle p \rangle-m_{0}$, respectively. On the assumption of
isotropic emission in the source rest frame, there is $\langle u
\rangle=(\pi/2)\langle u_{T} \rangle$. Besides, $\langle u_{T}
\rangle$ and $k^{'}$ from corresponding $ T -{\overline{m}}$
correlation approximatively meet $k^{'}=(1/2)\langle u_{T} \rangle
^{2}$, which means $\langle u_{T} \rangle$ and $\langle u \rangle$
can be obtained from correlation $ T -{\overline{m}}$ based on
$k^{'}=(1/2)\langle u_{T} \rangle ^{2}$ and $\langle u
\rangle=(\pi/2)\langle u_{T} \rangle$ relationships. In addition,
the mean transverse flow velocity and mean flow velocity have a
slightly increase tendency with center-of-mass energy, event
centrality, and system size.
\\

{\bf Conflict of Interests}

The authors declare that there is no conflict of interests
regarding the publication of this paper.
\\

{\bf Acknowledgments}

This work was supported by the National Natural Science Foundation
of China under Grant No. 11575103 and the US DOE under contract
DE-FG02-87ER40331.A008.

\renewcommand{\baselinestretch}{0.6}

\newpage
{\tiny {Table 1. Values of free parameters, normalization
constant,
 and $\chi^2$/dof corresponding to one- or two-component Erlang
 distribution in Figures 1--5 except Figure 3(a). The value
of $\chi^2$/dof for $\Xi^{-}+\bar{\Xi}^{+}$ in Figure 1(c) is in
fact the value of $\chi^2$ due to less data points.
{%
\begin{center}
\begin{tabular}{ccccccccc}
\hline\hline Figure & Type & $m_{1}$ & $p_{ti1}$ (GeV/$c^2$) & $k_{1}$  & $m_{2}$ & $p_{ti2}$ (GeV/$c^2$) & $N_{\rm E0}$ & $\chi^2$/dof \\
\hline
Figure 1(a) & $\pi^{+}$               & 2 & $0.175\pm0.010$ & 1 & - & - & $1.242\pm0.164$  & 0.032 \\
            & $K^{+}$                 & 2 & $0.280\pm0.030$ & 1 & - & - & $0.093\pm0.014$  & 0.009 \\
            & $p$                     & 2 & $0.330\pm0.030$ & 1 & - & - & $0.071\pm0.010$  & 0.027 \\
Figure 1(b) & $\pi^+$                 & 3 & $0.095\pm0.010$ & $0.515\pm0.050$ & 2 & $0.272\pm0.015$ & $1.840\pm0.200$ & 0.018 \\
            & $K^+$                   & 3 & $0.159\pm0.010$ & $0.670\pm0.050$ & 4 & $0.256\pm0.050$ & $0.230\pm0.030$ & 0.074 \\
            & $p$                     & 4 & $0.150\pm0.010$ & $0.550\pm0.050$ & 4 & $0.278\pm0.020$ & $0.101\pm0.010$ & 0.016 \\
Figure 1(c) & $\Lambda$               & 2 & $0.405\pm0.020$ & 1 & - & - & $0.054\pm0.023$  & 0.310 \\
            & $\phi$                  & 2 & $0.458\pm0.020$ & 1 & - & - & $0.020\pm0.022$  & 0.622 \\
            & $\Xi^{-}+\bar{\Xi}^{+}$ & 2 & $0.420\pm0.030$ & 1 & - & - & $0.007\pm0.030$  & (0.218) \\
Figure 1(d) & $\pi^+$                 & 3 & $0.096\pm0.010$ & $0.520\pm0.050$ & 2 & $0.300\pm0.020$ & $2.310\pm0.200$ & 0.025 \\
            & $K^+$                   & 3 & $0.162\pm0.010$ & $0.620\pm0.050$ & 4 & $0.250\pm0.030$ & $0.294\pm0.020$ & 0.015 \\
            & $p$                     & 3 & $0.245\pm0.020$ & $0.790\pm0.070$ & 4 & $0.355\pm0.080$ & $0.132\pm0.010$ & 0.057 \\
Figure 1(e) & $\pi^+$                 & 3 & $0.096\pm0.010$ & $0.515\pm0.060$ & 2 & $0.310\pm0.030$ & $2.850\pm0.300$ & 0.099 \\
            & $K^+$                   & 3 & $0.167\pm0.010$ & $0.620\pm0.080$ & 4 & $0.260\pm0.030$ & $0.360\pm0.030$ & 0.015 \\
            & $p$                     & 3 & $0.242\pm0.020$ & $0.670\pm0.070$ & 4 & $0.357\pm0.070$ & $0.168\pm0.015$ & 0.017 \\
Figure 1(f) & $\Xi$                   & 3 & $0.400\pm0.040$ & $0.870\pm0.050$ & 2 & $0.920\pm0.060$ & $(2.320\pm0.500) \times 10^{-3}$ & 0.056 \\
            & $\Omega$                & 2 & $0.650\pm0.040$ & $0.980\pm0.020$ & 2 & $1.100\pm0.050$ & $(2.480\pm0.600) \times 10^{-4}$ & 0.064 \\
\hline
Figure 2(a) & 0-10\%                  & 2 & $0.310\pm0.030$ & $0.953\pm0.020$ & 1 & $0.760\pm0.040$ & $6.426\pm2.734$ & 0.301 \\
            & 20-30\%                 & 2 & $0.310\pm0.040$ & $0.940\pm0.030$ & 1 & $0.740\pm0.040$ & $3.011\pm1.369$ & 0.214 \\
            & 40-60\%                 & 2 & $0.300\pm0.040$ & $0.920\pm0.030$ & 1 & $0.730\pm0.050$ & $1.067\pm0.421$ & 0.368 \\
Figure 2(b) & 0-10\%                  & 4 & $0.275\pm0.020$ & $0.988\pm0.008$ & 3 & $0.530\pm0.040$ & $0.822\pm0.363$ & 0.296 \\
            & 20-30\%                 & 4 & $0.260\pm0.020$ & $0.972\pm0.015$ & 3 & $0.500\pm0.040$ & $0.446\pm0.166$ & 0.088 \\
            & 40-60\%                 & 4 & $0.247\pm0.020$ & $0.957\pm0.020$ & 3 & $0.485\pm0.040$ & $0.154\pm0.067$ & 0.233 \\
Figure 2(c) & 0-10\%                  & 3 & $0.360\pm0.030$ & 1 & - & - & $0.131\pm0.055$  & 0.655 \\
            & 20-30\%                 & 3 & $0.370\pm0.030$ & 1 & - & - & $0.051\pm0.020$  & 1.353 \\
            & 40-60\%                 & 3 & $0.363\pm0.030$ & 1 & - & - & $0.015\pm0.008$  & 0.708 \\
Figure 2(d) & 0-10\%                  & 4 & $0.362\pm0.020$ & 1 & - & - & $0.018\pm0.006$  & 0.419 \\
            & 20-30\%                 & 3 & $0.393\pm0.025$ & 1 & - & - & $0.014\pm0.005$  & 0.139 \\
            & 40-60\%                 & 3 & $0.400\pm0.030$ & 1 & - & - & $0.003\pm0.001$  & 0.121 \\
\hline
Figure 3(b) & 0-5\%                   & 2 & $0.330\pm0.030$ & $0.982\pm0.006$ & 1 & $0.840\pm0.050$ & $18.680\pm5.189$ & 0.034 \\
            & 10-20\%                 & 2 & $0.337\pm0.030$ & $0.960\pm0.010$ & 1 & $0.780\pm0.050$ & $10.605\pm2.555$ & 0.028 \\
            & 20-40\%                 & 2 & $0.330\pm0.030$ & $0.950\pm0.020$ & 1 & $0.760\pm0.040$ & $6.766\pm2.082$ & 0.015 \\
            & 40-60\%                 & 2 & $0.326\pm0.030$ & $0.957\pm0.020$ & 1 & $0.810\pm0.050$ & $2.623\pm1.049$ & 0.056 \\
            & 60-80\%                 & 2 & $0.310\pm0.030$ & $0.900\pm0.030$ & 1 & $0.710\pm0.050$ & $0.879\pm0.274$ & 0.042 \\
Figure 3(c) & 0-12\%                  & 4 & $0.270\pm0.020$ & $0.986\pm0.005$ & 1 & $1.030\pm0.050$ & $5.061\pm1.250$ & 0.254 \\
            & 10-20\%                 & 4 & $0.260\pm0.020$ & $0.920\pm0.020$ & 1 & $0.810\pm0.050$ & $4.106\pm1.392$ & 0.136 \\
            & 20-40\%                 & 4 & $0.255\pm0.030$ & $0.920\pm0.040$ & 1 & $0.800\pm0.050$ & $2.408\pm0.992$ & 0.188 \\
            & 40-60\%                 & 4 & $0.237\pm0.030$ & $0.870\pm0.040$ & 1 & $0.750\pm0.050$ & $1.299\pm0.475$ & 0.195 \\
            & 60-80\%                 & 4 & $0.210\pm0.030$ & $0.800\pm0.050$ & 1 & $0.680\pm0.050$ & $0.482\pm0.204$ & 0.117 \\
Figure 3(d) & 0-5\%                   & 3 & $0.340\pm0.020$ & 1 & - & - & $1.762\pm0.587$  & 0.019 \\
            & 10-20\%                 & 3 & $0.355\pm0.020$ & 1 & - & - & $1.055\pm0.352$  & 0.031 \\
            & 30-40\%                 & 3 & $0.355\pm0.020$ & 1 & - & - & $0.478\pm0.154$  & 0.026 \\
            & 50-60\%                 & 2 & $0.445\pm0.035$ & 1 & - & - & $0.276\pm0.085$  & 0.071 \\
            & 70-80\%                 & 2 & $0.415\pm0.040$ & 1 & - & - & $0.068\pm0.023$  & 0.110 \\
Figure 3(e) & 0-5\%                   & 4 & $0.286\pm0.015$ & $0.998\pm0.001$ & 3 & $0.640\pm0.040$ & $2.970\pm1.165$ & 0.035 \\
            & 10-20\%                 & 4 & $0.286\pm0.015$ & $0.996\pm0.002$ & 3 & $0.590\pm0.040$ & $1.839\pm0.582$ & 0.049 \\
            & 20-40\%                 & 4 & $0.281\pm0.020$ & $0.991\pm0.004$ & 3 & $0.590\pm0.040$ & $1.077\pm0.414$ & 0.069 \\
            & 40-60\%                 & 4 & $0.270\pm0.020$ & $0.977\pm0.010$ & 3 & $0.530\pm0.040$ & $0.466\pm0.159$ & 0.009 \\
            & 60-80\%                 & 4 & $0.250\pm0.025$ & $0.950\pm0.020$ & 3 & $0.470\pm0.040$ & $0.145\pm0.053$ & 0.059 \\
Figure 3(f) & 0-5\%                   & 4 & $0.324\pm0.020$ & 1 & - & - & $0.270\pm0.090$  & 0.008 \\
            & 10-20\%                 & 4 & $0.314\pm0.020$ & 1 & - & - & $0.190\pm0.064$  & 0.044 \\
            & 20-40\%                 & 4 & $0.315\pm0.020$ & 1 & - & - & $0.098\pm0.030$  & 0.052 \\
            & 40-60\%                 & 4 & $0.309\pm0.030$ & 1 & - & - & $0.038\pm0.013$  & 0.136 \\
            & 60-80\%                 & 4 & $0.325\pm0.040$ & 1 & - & - & $0.008\pm0.003$  & 0.402 \\
\hline
Figure 4(a) & 0-5\%                   & 2 & $0.234\pm0.020$ & $0.580\pm0.080$ & 1 & $0.530\pm0.030$ & $512.268\pm123.936$ & 0.058 \\
            & 50-60\%                 & 2 & $0.194\pm0.030$ & $0.520\pm0.020$ & 1 & $0.530\pm0.030$ & $54.364\pm13.591$   & 0.038 \\
            & 80-90\%                 & 2 & $0.170\pm0.030$ & $0.550\pm0.050$ & 1 & $0.505\pm0.035$ & $5.474\pm1.466$     & 0.037 \\
Figure 4(b) & 0-5\%                   & 3 & $0.280\pm0.030$ & $0.850\pm0.050$ & 2 & $0.600\pm0.050$ & $30.686\pm8.768$    & 0.025 \\
            & 50-60\%                 & 3 & $0.220\pm0.040$ & $0.550\pm0.050$ & 2 & $0.480\pm0.050$ & $3.415\pm1.067$     & 0.022 \\
            & 80-90\%                 & 3 & $0.190\pm0.050$ & $0.550\pm0.050$ & 2 & $0.460\pm0.050$ & $0.307\pm0.095$     & 0.054 \\
Figure 4(c) & 0-5\%                   & 3 & $0.430\pm0.030$ & 1 & - & - & $5.865\pm2.091$  & 0.275 \\
            & 50-60\%                 & 3 & $0.370\pm0.030$ & $0.950\pm0.030$ & 3 & $0.600\pm0.070$ & $0.741\pm0.265$     & 0.041 \\
            & 80-90\%                 & 3 & $0.285\pm0.040$ & $0.900\pm0.050$ & 3 & $0.520\pm0.080$ & $0.094\pm0.033$     & 0.053 \\
Figure 4(d) & 0-5\%                   & 3 & $0.460\pm0.030$ & 1 & - & - & $12.499\pm3.000$ & 0.289 \\
            & 50-60\%                 & 2 & $0.580\pm0.040$ & 1 & - & - & $1.569\pm0.500$  & 0.272 \\
            & 80-90\%                 & 2 & $0.570\pm0.040$ & 1 & - & - & $0.105\pm0.035$  & 0.344 \\
\hline
Figure 5(a) & 0-5\%                   & 2 & $0.175\pm0.040$ & $0.680\pm0.050$ & 2 & $0.460\pm0.030$ & $28.109\pm8.031$ & 0.176 \\
            & 40-60\%                 & 2 & $0.165\pm0.040$ & $0.710\pm0.050$ & 2 & $0.440\pm0.030$ & $10.766\pm3.014$ & 0.083 \\
            & 80-100\%                & 2 & $0.175\pm0.040$ & $0.830\pm0.050$ & 2 & $0.440\pm0.030$ & $3.124\pm0.892$  & 0.094 \\
Figure 5(b) & 0-5\%                   & 3 & $0.235\pm0.030$ & $0.520\pm0.100$ & 2 & $0.600\pm0.050$ & $1.738\pm0.280$  & 0.020 \\
            & 40-60\%                 & 3 & $0.210\pm0.030$ & $0.550\pm0.100$ & 2 & $0.550\pm0.050$ & $0.671\pm0.104$  & 0.031 \\
            & 80-100\%                & 3 & $0.165\pm0.030$ & $0.520\pm0.100$ & 2 & $0.440\pm0.040$ & $0.218\pm0.039$  & 0.011 \\
Figure 5(c) & 0-5\%                   & 3 & $0.400\pm0.030$ & $0.560\pm0.100$ & 2 & $0.650\pm0.050$ & $0.450\pm0.108$  & 0.036 \\
            & 40-60\%                 & 3 & $0.285\pm0.040$ & $0.680\pm0.100$ & 3 & $0.510\pm0.030$ & $0.203\pm0.045$  & 0.012 \\
            & 80-100\%                & 3 & $0.237\pm0.030$ & $0.820\pm0.050$ & 3 & $0.465\pm0.030$ & $0.069\pm0.015$  & 0.022 \\
Figure 5(d) & 0-5\%                   & 4 & $0.345\pm0.040$ & $0.710\pm0.070$ & 2 & $0.820\pm0.040$ & $0.256\pm0.051$  & 0.026 \\
            & 40-60\%                 & 4 & $0.285\pm0.040$ & $0.740\pm0.060$ & 2 & $0.780\pm0.040$ & $0.103\pm0.024$  & 0.082 \\
            & 80-100\%                & 3 & $0.313\pm0.040$ & $0.880\pm0.040$ & 2 & $0.780\pm0.040$ & $0.031\pm0.009$  & 0.074 \\
\hline\hline
\end{tabular}%
\end{center}
}} }

\newpage
{\tiny {Table 2. Values of free parameters, normalization
constants, and $\chi^2$/dof corresponding to Tsallis distribution
in Figures 1--5. The value of $\chi^2$/dof for
$\Xi^{-}+\bar{\Xi}^{+}$ in Figure 1(c) is the value of $\chi^2$
due to less data points.
{%
\begin{center}
\begin{tabular}{cccccc}
\hline\hline  Figure & Type & $T$ (GeV)& $q$& $N_{\rm T0}$ & $\chi^2$/dof \\
\hline
Figure 1(a) & $\pi^{+}$              & $0.088\pm0.002$ & $1.095\pm0.010$ & $1.048\pm0.113$          & 0.018 \\
            & $K^{+}$                & $0.105\pm0.003$ & $1.074\pm0.010$ & $0.081\pm0.012$          & 0.037 \\
            & $p$                    & $0.135\pm0.003$ & $1.045\pm0.008$ & $0.051\pm0.006$          & 0.085 \\
Figure 1(b) & $\pi^+$                & $0.092\pm0.003$ & $1.127\pm0.007$ & $1.852\pm0.308$          & 0.077 \\
            & $K^+$                  & $0.122\pm0.004$ & $1.107\pm0.007$ & $0.229\pm0.029$          & 0.017 \\
            & $p$                    & $0.159\pm0.004$ & $1.069\pm0.007$ & $0.103\pm0.015$          & 0.084 \\
Figure 1(c) & $\Lambda$              & $0.175\pm0.003$ & $1.064\pm0.002$ & $0.045\pm0.005$          & 1.192 \\
            & $\phi$                 & $0.174\pm0.003$ & $1.092\pm0.002$ & $0.018\pm0.002$          & 0.804 \\
            & $\Xi^{-}+\bar{\Xi}^{+}$& $0.180\pm0.003$ & $1.060\pm0.004$ & $0.012\pm0.002$          & (0.348) \\
Figure 1(d) & $\pi^+$                & $0.092\pm0.003$ & $1.140\pm0.007$ & $2.305\pm0.360$          & 0.131 \\
            & $K^+$                  & $0.123\pm0.003$ & $1.121\pm0.008$ & $0.302\pm0.040$          & 0.014 \\
            & $p$                    & $0.159\pm0.003$ & $1.082\pm0.007$ & $0.134\pm0.021$          & 0.081 \\
Figure 1(e) & $\pi^+$                & $0.090\pm0.003$ & $1.150\pm0.005$ & $2.920\pm0.319$          & 0.143 \\
            & $K^+$                  & $0.127\pm0.003$ & $1.124\pm0.005$ & $0.376\pm0.036$          & 0.022 \\
            & $p$                    & $0.163\pm0.003$ & $1.098\pm0.005$ & $0.168\pm0.016$          & 0.042 \\
Figure 1(f) & $\Xi$                  & $0.188\pm0.003$ & $1.099\pm0.002$ & $(2.389\pm0.342) \times 10^{-3}$& 0.583 \\
            & $\Omega$               & $0.215\pm0.003$ & $1.096\pm0.003$ & $(2.180\pm0.322) \times 10^{-4}$& 0.037 \\
\hline
Figure 2(a) & 0-10\%                 & $0.132\pm0.007$ & $1.085\pm0.004$ & $7.155\pm2.600$          & 0.242 \\
            & 20-30\%                & $0.128\pm0.007$ & $1.087\pm0.004$ & $3.504\pm1.460$          & 0.290 \\
            & 40-60\%                & $0.125\pm0.007$ & $1.092\pm0.004$ & $1.148\pm0.546$          & 0.195 \\
Figure 2(b) & 0-10\%                 & $0.227\pm0.007$ & $1.034\pm0.004$ & $1.620\pm0.733$          & 0.789 \\
            & 20-30\%                & $0.198\pm0.007$ & $1.045\pm0.004$ & $0.801\pm0.280$          & 0.275 \\
            & 40-60\%                & $0.172\pm0.006$ & $1.055\pm0.004$ & $0.270\pm0.104$          & 0.331 \\
Figure 2(c) & 0-10\%                 & $0.243\pm0.006$ & $1.033\pm0.003$ & $0.167\pm0.048$          & 0.167 \\
            & 20-30\%                & $0.228\pm0.006$ & $1.042\pm0.004$ & $0.065\pm0.023$          & 0.379 \\
            & 40-60\%                & $0.210\pm0.006$ & $1.043\pm0.003$ & $0.022\pm0.008$          & 0.343 \\
Figure 2(d) & 0-10\%                 & $0.306\pm0.007$ & $1.029\pm0.005$ & $0.032\pm0.009$          & 0.858 \\
            & 20-30\%                & $0.233\pm0.007$ & $1.044\pm0.004$ & $0.016\pm0.004$          & 0.089 \\
            & 40-60\%                & $0.223\pm0.007$ & $1.048\pm0.004$ & $0.003\pm0.001$          & 0.080 \\
\hline
Figure 3(a) & 0-12\%                 & $0.112\pm0.006$ & $1.095\pm0.005$ & $168.987\pm77.994$       & 0.115 \\
            & 10-20\%                & $0.110\pm0.005$ & $1.097\pm0.005$ & $137.542\pm75.023$       & 0.124 \\
            & 20-40\%                & $0.109\pm0.006$ & $1.100\pm0.005$ & $73.740\pm36.870$        & 0.058 \\
            & 40-60\%                & $0.106\pm0.006$ & $1.103\pm0.004$ & $30.072\pm16.193$        & 0.028 \\
            & 60-80\%                & $0.103\pm0.006$ & $1.105\pm0.005$ & $9.980\pm5.420$          & 0.084 \\
Figure 3(b) & 0-5\%                  & $0.166\pm0.006$ & $1.069\pm0.005$ & $18.141\pm6.977$         & 0.142 \\
            & 10-20\%                & $0.163\pm0.006$ & $1.073\pm0.005$ & $11.295\pm4.651$         & 0.056 \\
            & 20-40\%                & $0.162\pm0.006$ & $1.074\pm0.005$ & $6.527\pm2.611$          & 0.032 \\
            & 40-60\%                & $0.159\pm0.006$ & $1.078\pm0.005$ & $2.607\pm1.117$          & 0.049 \\
            & 60-80\%                & $0.157\pm0.007$ & $1.078\pm0.005$ & $0.772\pm0.332$          & 0.046 \\
Figure 3(c) & 0-12\%                 & $0.215\pm0.007$ & $1.055\pm0.005$ & $7.233\pm3.945$          & 1.785 \\
            & 10-20\%                & $0.210\pm0.010$ & $1.054\pm0.006$ & $5.883\pm2.942$          & 0.442 \\
            & 20-40\%                & $0.203\pm0.010$ & $1.055\pm0.005$ & $3.389\pm1.694$          & 0.562 \\
            & 40-60\%                & $0.185\pm0.010$ & $1.062\pm0.005$ & $1.452\pm0.581$          & 0.296 \\
            & 60-80\%                & $0.160\pm0.010$ & $1.071\pm0.005$ & $0.466\pm0.250$          & 0.283 \\
Figure 3(d) & 0-5\%                  & $0.255\pm0.015$ & $1.033\pm0.010$ & $2.107\pm0.892$          & 0.041 \\
            & 10-20\%                & $0.253\pm0.013$ & $1.037\pm0.012$ & $1.381\pm0.642$          & 0.066 \\
            & 30-40\%                & $0.250\pm0.015$ & $1.038\pm0.010$ & $0.613\pm0.245$          & 0.079 \\
            & 50-60\%                & $0.222\pm0.012$ & $1.055\pm0.010$ & $0.212\pm0.091$          & 0.061 \\
            & 70-80\%                & $0.203\pm0.015$ & $1.052\pm0.010$ & $0.054\pm0.024$          & 0.105 \\
Figure 3(e) & 0-5\%                  & $0.288\pm0.015$ & $1.018\pm0.005$ & $4.465\pm2.436$          & 0.378 \\
            & 10-20\%                & $0.278\pm0.013$ & $1.022\pm0.006$ & $2.574\pm1.065$          & 0.161 \\
            & 20-40\%                & $0.273\pm0.011$ & $1.027\pm0.007$ & $1.436\pm0.676$          & 0.370 \\
            & 40-60\%                & $0.260\pm0.012$ & $1.030\pm0.006$ & $0.620\pm0.275$          & 0.161 \\
            & 60-80\%                & $0.250\pm0.013$ & $1.030\pm0.006$ & $0.204\pm0.102$          & 0.198 \\
Figure 3(f) & 0-5\%                  & $0.365\pm0.015$ & $1.0011\pm0.001$& $0.415\pm0.183$          & 0.061 \\
            & 10-20\%                & $0.352\pm0.015$ & $1.0011\pm0.001$& $0.301\pm0.113$          & 0.059 \\
            & 20-40\%                & $0.345\pm0.015$ & $1.0011\pm0.001$& $0.171\pm0.068$          & 0.034 \\
            & 40-60\%                & $0.332\pm0.015$ & $1.0011\pm0.001$& $0.068\pm0.034$          & 0.140 \\
            & 60-80\%                & $0.328\pm0.015$ & $1.011\pm0.010$ & $0.015\pm0.006$          & 0.248 \\
\hline
Figure 4(a) & 0-5\%                  & $0.112\pm0.002$ & $1.126\pm0.005$ & $430.964\pm121.590$      & 0.211 \\
            & 50-60\%                & $0.091\pm0.002$ & $1.147\pm0.005$ & $48.274\pm10.284$        & 0.161 \\
            & 80-90\%                & $0.071\pm0.002$ & $1.163\pm0.005$ & $5.183\pm0.952$          & 0.114 \\
Figure 4(b) & 0-5\%                  & $0.258\pm0.005$ & $1.038\pm0.005$ & $34.752\pm5.111$         & 0.009 \\
            & 50-60\%                & $0.160\pm0.004$ & $1.107\pm0.005$ & $3.822\pm0.637$          & 0.019 \\
            & 80-90\%                & $0.125\pm0.004$ & $1.123\pm0.005$ & $0.346\pm0.077$          & 0.048 \\
Figure 4(c) & 0-5\%                  & $0.380\pm0.010$ & $1.023\pm0.010$ & $7.346\pm1.920$          & 0.362 \\
            & 50-60\%                & $0.305\pm0.006$ & $1.037\pm0.005$ & $0.874\pm0.204$          & 0.041 \\
            & 80-90\%                & $0.200\pm0.005$ & $1.064\pm0.005$ & $0.104\pm0.025$          & 0.104 \\
Figure 4(d) & 0-5\%                  & $0.379\pm0.007$ & $1.026\pm0.005$ & $13.487\pm2.810$         & 0.223 \\
            & 50-60\%                & $0.350\pm0.007$ & $1.040\pm0.007$ & $1.400\pm0.262$          & 0.302 \\
            & 80-90\%                & $0.256\pm0.007$ & $1.075\pm0.007$ & $0.096\pm0.021$          & 0.182 \\
\hline
Figure 5(a) & 0-5\%                  & $0.099\pm0.003$ & $1.163\pm0.005$ & $23.932\pm5.400$         & 0.371 \\
            & 40-60\%                & $0.088\pm0.003$ & $1.163\pm0.006$ & $10.008\pm2.045$         & 0.304 \\
            & 80-100\%               & $0.072\pm0.003$ & $1.164\pm0.006$ & $3.338\pm0.747$          & 0.014 \\
Figure 5(b) & 0-5\%                  & $0.177\pm0.003$ & $1.129\pm0.005$ & $1.915\pm0.231$          & 0.007 \\
            & 40-60\%                & $0.144\pm0.003$ & $1.135\pm0.005$ & $0.750\pm0.094$          & 0.010 \\
            & 80-100\%               & $0.090\pm0.002$ & $1.152\pm0.005$ & $0.235\pm0.033$          & 0.031 \\
Figure 5(c) & 0-5\%                  & $0.295\pm0.009$ & $1.068\pm0.008$ & $0.515\pm0.124$          & 0.102 \\
            & 40-60\%                & $0.184\pm0.006$ & $1.100\pm0.007$ & $0.237\pm0.057$          & 0.032 \\
            & 80-100\%               & $0.091\pm0.004$ & $1.120\pm0.005$ & $0.079\pm0.022$          & 0.018 \\
Figure 5(d) & 0-5\%                  & $0.301\pm0.005$ & $1.072\pm0.004$ & $0.348\pm0.083$          & 0.025 \\
            & 40-60\%                & $0.196\pm0.005$ & $1.098\pm0.004$ & $0.144\pm0.032$          & 0.018 \\
            & 80-100\%               & $0.100\pm0.003$ & $1.119\pm0.003$ & $0.040\pm0.013$          & 0.026 \\
\hline\hline
\end{tabular}%
\end{center}
}} }

\vskip1.0cm

{\tiny {Table 3. Values of free parameters, normalization
constant, and $\chi^2$/dof corresponding to the three-component
Erlang distribution in Figure 3(a), where $m_1=m_2=m_3=2$ which
are not listed in the column.
{%
\begin{center}
\begin{tabular}{ccccccccc}
\hline\hline Figure & Type & $p_{ti1}$ (GeV/$c^2$) & $k_{1}$ & $p_{ti2}$ (GeV/$c^2$) & $k_{2}$ & $p_{ti3}$ (GeV/$c^2$) & $N_{\rm E0}$ & $\chi^2$/dof \\
\hline
Figure 3(a) & 0-12\%  & $0.220\pm0.040$ & $0.90971\pm0.03640$ & $0.400\pm0.040$ & $0.08997\pm0.00360$ & $1.040\pm0.050$& $131.650\pm62.361$  & 0.043 \\
            & 10-20\% & $0.221\pm0.040$ & $0.90964\pm0.03640$ & $0.400\pm0.040$ & $0.08996\pm0.00361$ & $1.050\pm0.080$& $93.234\pm34.531$   & 0.119 \\
            & 20-40\% & $0.215\pm0.040$ & $0.87930\pm0.04400$ & $0.390\pm0.040$ & $0.11990\pm0.00600$ & $0.960\pm0.060$& $52.239\pm24.378$   & 0.086 \\
            & 40-60\% & $0.212\pm0.040$ & $0.90936\pm0.04550$ & $0.415\pm0.040$ & $0.08994\pm0.00451$ & $1.000\pm0.070$& $24.136\pm7.099$    & 0.080 \\
            & 60-80\% & $0.200\pm0.040$ & $0.90936\pm0.03640$ & $0.420\pm0.040$ & $0.08994\pm0.00361$ & $1.010\pm0.080$& $8.128\pm3.695$     & 0.183 \\
\hline
\end{tabular}%
\end{center}
}} }

\newpage
{\tiny {Table 4. Values of intercepts, slopes, and $\chi^2$
corresponding to the lines in Figures 6--11.
{%
\begin{center}
\begin{tabular}{cccccc}
\hline\hline  Figure & Correlation & Type & Intercept & Slope  & $\chi^2$ \\
\hline
Figure 6(a) &$\langle p_{T}\rangle-m_{0}$ & $p$-$p$ 0.2 TeV   & $0.322\pm0.064$ & $0.386\pm0.104$ & 0.175 \\
            &                             & $p$-$p$ 0.9 TeV   & $0.425\pm0.081$ & $0.382\pm0.087$ & 1.001 \\
            &                             & $p$-$p$ 2.76 TeV  & $0.377\pm0.041$ & $0.554\pm0.067$ & 0.050 \\
            &                             & $p$-$p$ 7 TeV     & $0.398\pm0.064$ & $0.597\pm0.060$ & 0.187 \\
Figure 6(b) &                             & 0-10\%            & $0.295\pm0.126$ & $0.669\pm0.102$ & 0.154 \\
            &                             & 20-30\%           & $0.433\pm0.120$ & $0.486\pm0.098$ & 0.184 \\
            &                             & 40-60\%           & $0.386\pm0.078$ & $0.513\pm0.064$ & 0.077 \\
Figure 6(c) &                             & 0-12\%            & $0.353\pm0.047$ & $0.708\pm0.050$ & 0.160 \\
            &                             & 10-20\%           & $0.367\pm0.020$ & $0.684\pm0.022$ & 0.036 \\
            &                             & 20-40\%           & $0.358\pm0.020$ & $0.686\pm0.021$ & 0.043 \\
            &                             & 40-60\%           & $0.347\pm0.059$ & $0.632\pm0.064$ & 0.209 \\
            &                             & 60-80\%           & $0.296\pm0.105$ & $0.647\pm0.114$ & 0.524 \\
Figure 6(d) &                             & 0-5\%             & $0.372\pm0.027$ & $0.994\pm0.036$ & 0.037 \\
            &                             & 50-60\%           & $0.362\pm0.040$ & $0.815\pm0.054$ & 0.057 \\
            &                             & 80-90\%           & $0.325\pm0.084$ & $0.738\pm0.114$ & 0.355 \\
Figure 6(e) &                             & 0-5\%             & $0.436\pm0.054$ & $0.906\pm0.070$ & 0.088 \\
            &                             & 40-60\%           & $0.417\pm0.054$ & $0.740\pm0.070$ & 0.093 \\
            &                             & 80-100\%          & $0.376\pm0.049$ & $0.546\pm0.063$ & 0.071 \\
\hline
Figure 7(a) &$\langle p \rangle-m_{0}$    & $p$-$p$ 0.2 TeV   & $0.506\pm0.010$ & $0.606\pm0.163$ & 0.336 \\
            &                             & $p$-$p$ 0.9 TeV   & $0.666\pm0.127$ & $0.600\pm0.137$ & 2.176 \\
            &                             & $p$-$p$ 2.76 TeV  & $0.592\pm0.065$ & $0.870\pm0.105$ & 0.091 \\
            &                             & $p$-$p$ 7 TeV     & $0.626\pm0.101$ & $0.937\pm0.095$ & 0.401 \\
Figure 7(b) &                             & 0-10\%            & $0.463\pm0.197$ & $1.051\pm0.161$ & 0.369 \\
            &                             & 20-30\%           & $0.680\pm0.189$ & $0.763\pm0.154$ & 0.430 \\
            &                             & 40-60\%           & $0.606\pm0.123$ & $0.806\pm0.100$ & 0.181 \\
Figure 7(c) &                             & 0-12\%            & $0.554\pm0.074$ & $1.112\pm0.079$ & 0.327 \\
            &                             & 10-20\%           & $0.576\pm0.032$ & $1.075\pm0.034$ & 0.071 \\
            &                             & 20-40\%           & $0.562\pm0.031$ & $1.077\pm0.033$ & 0.081 \\
            &                             & 40-60\%           & $0.545\pm0.093$ & $0.993\pm0.100$ & 0.482 \\
            &                             & 60-80\%           & $0.465\pm0.165$ & $1.016\pm0.178$ & 1.182 \\
Figure 7(d) &                             & 0-5\%             & $0.585\pm0.042$ & $1.561\pm0.057$ & 0.062 \\
            &                             & 50-60\%           & $0.569\pm0.063$ & $1.279\pm0.086$ & 0.104 \\
            &                             & 80-90\%           & $0.510\pm0.132$ & $1.159\pm0.180$ & 0.300 \\
Figure 7(e) &                             & 0-5\%             & $0.685\pm0.086$ & $1.424\pm0.111$ & 0.155 \\
            &                             & 40-60\%           & $0.655\pm0.085$ & $1.162\pm0.110$ & 0.168 \\
            &                             & 80-100\%          & $0.590\pm0.076$ & $0.858\pm0.099$ & 0.155 \\
\hline
Figure 8(a) &$T-m_{0}$                    & $p$-$p$ 0.2 TeV   & $0.078\pm0.004$ & $0.060\pm0.006$ & 0.038 \\
            &                             & $p$-$p$ 0.9 TeV   & $0.083\pm0.006$ & $0.080\pm0.007$ & 0.382 \\
            &                             & $p$-$p$ 2.76 TeV  & $0.080\pm0.001$ & $0.080\pm0.001$ & 0.046 \\
            &                             & $p$-$p$ 7 TeV     & $0.084\pm0.004$ & $0.080\pm0.004$ & 0.186 \\
Figure 8(b) &                             & 0-10\%            & $0.059\pm0.011$ & $0.145\pm0.009$ & 0.073 \\
            &                             & 20-30\%           & $0.087\pm0.022$ & $0.095\pm0.018$ & 0.342 \\
            &                             & 40-60\%           & $0.082\pm0.016$ & $0.087\pm0.013$ & 0.193 \\
Figure 8(c) &                             & 0-12\%            & $0.068\pm0.026$ & $0.197\pm0.028$ & 1.545 \\
            &                             & 10-20\%           & $0.069\pm0.024$ & $0.190\pm0.026$ & 1.204 \\
            &                             & 20-40\%           & $0.069\pm0.025$ & $0.184\pm0.027$ & 1.174 \\
            &                             & 40-60\%           & $0.068\pm0.030$ & $0.170\pm0.032$ & 1.642 \\
            &                             & 60-80\%           & $0.064\pm0.039$ & $0.162\pm0.042$ & 2.843 \\
Figure 8(d) &                             & 0-5\%             & $0.083\pm0.025$ & $0.308\pm0.034$ & 1.696 \\
            &                             & 50-60\%           & $0.036\pm0.022$ & $0.294\pm0.029$ & 1.922 \\
            &                             & 80-90\%           & $0.037\pm0.022$ & $0.195\pm0.029$ & 1.057 \\
Figure 8(e) &                             & 0-5\%             & $0.070\pm0.013$ & $0.220\pm0.018$ & 0.319 \\
            &                             & 40-60\%           & $0.083\pm0.009$ & $0.109\pm0.012$ & 0.558 \\
            &                             & 80-100\%          & $0.074\pm0.005$ & $0.023\pm0.007$ & 0.296 \\
\hline
Figure 9(a) &$\langle p_{T}\rangle-{\overline{m}}$ & $p$-$p$ 0.2 TeV& $0.167\pm0.056$ & $0.339\pm0.063$ & 0.062 \\
            &                             & $p$-$p$ 0.9 TeV   & $0.207\pm0.046$ & $0.346\pm0.075$ & 0.324 \\
            &                             & $p$-$p$ 2.76 TeV  & $0.148\pm0.028$ & $0.418\pm0.038$ & 0.015 \\
            &                             & $p$-$p$ 7 TeV     & $0.160\pm0.023$ & $0.421\pm0.047$ & 0.061 \\
Figure 9(b) &                             & 0-10\%            & $0.097\pm0.032$ & $0.451\pm0.071$ & 0.035 \\
            &                             & 20-30\%           & $0.199\pm0.044$ & $0.385\pm0.094$ & 0.060 \\
            &                             & 40-60\%           & $0.178\pm0.030$ & $0.391\pm0.062$ & 0.028 \\
Figure 9(c) &                             & 0-12\%            & $0.105\pm0.013$ & $0.468\pm0.026$ & 0.025 \\
            &                             & 10-20\%           & $0.116\pm0.006$ & $0.460\pm0.012$ & 0.005 \\
            &                             & 20-40\%           & $0.113\pm0.005$ & $0.461\pm0.009$ & 0.005 \\
            &                             & 40-60\%           & $0.112\pm0.022$ & $0.448\pm0.039$ & 0.054 \\
            &                             & 60-80\%           & $0.076\pm0.034$ & $0.459\pm0.059$ & 0.129 \\
Figure 9(d) &                             & 0-5\%             & $0.081\pm0.007$ & $0.522\pm0.014$ & 0.005 \\
            &                             & 50-60\%           & $0.102\pm0.016$ & $0.486\pm0.029$ & 0.013 \\
            &                             & 80-90\%           & $0.097\pm0.024$ & $0.470\pm0.038$ & 0.016 \\
Figure 9(e) &                             & 0-5\%             & $0.101\pm0.012$ & $0.510\pm0.024$ & 0.008 \\
            &                             & 40-60\%           & $0.121\pm0.017$ & $0.475\pm0.030$ & 0.012 \\
            &                             & 80-100\%          & $0.147\pm0.022$ & $0.414\pm0.034$ & 0.017 \\
\hline
Figure 10(a)&$\langle p \rangle-{\overline{m}}$ & $p$-$p$ 0.2 TeV   & $0.262\pm0.099$ & $0.532\pm0.089$ & 0.122 \\
            &                             & $p$-$p$ 0.9 TeV   & $0.326\pm0.118$ & $0.544\pm0.072$ & 0.751 \\
            &                             & $p$-$p$ 2.76 TeV  & $0.232\pm0.059$ & $0.657\pm0.045$ & 0.028 \\
            &                             & $p$-$p$ 7 TeV     & $0.251\pm0.074$ & $0.661\pm0.037$ & 0.131 \\
Figure 10(b)&                             & 0-10\%            & $0.152\pm0.112$ & $0.708\pm0.050$ & 0.084 \\
            &                             & 20-30\%           & $0.312\pm0.148$ & $0.605\pm0.070$ & 0.145 \\
            &                             & 40-60\%           & $0.280\pm0.098$ & $0.615\pm0.046$ & 0.066 \\
Figure 10(c)&                             & 0-12\%            & $0.165\pm0.040$ & $0.736\pm0.021$ & 0.051 \\
            &                             & 10-20\%           & $0.183\pm0.018$ & $0.724\pm0.010$ & 0.011 \\
            &                             & 20-40\%           & $0.178\pm0.015$ & $0.724\pm0.008$ & 0.009 \\
            &                             & 40-60\%           & $0.175\pm0.062$ & $0.703\pm0.034$ & 0.126 \\
            &                             & 60-80\%           & $0.120\pm0.093$ & $0.721\pm0.053$ & 0.283 \\
Figure 10(d)&                             & 0-5\%             & $0.127\pm0.022$ & $0.820\pm0.011$ & 0.009 \\
            &                             & 50-60\%           & $0.160\pm0.045$ & $0.763\pm0.026$ & 0.025 \\
            &                             & 80-90\%           & $0.152\pm0.060$ & $0.738\pm0.037$ & 0.034 \\
Figure 10(e)&                             & 0-5\%             & $0.158\pm0.038$ & $0.800\pm0.019$ & 0.014 \\
            &                             & 40-60\%           & $0.190\pm0.047$ & $0.747\pm0.027$ & 0.023 \\
            &                             & 80-100\%          & $0.231\pm0.054$ & $0.651\pm0.035$ & 0.037 \\
\hline
Figure 11(a)&$T-{\overline{m}}$           & $p$-$p$ 0.2 TeV   & $0.056\pm0.011$ & $0.051\pm0.010$ & 0.143 \\
            &                             & $p$-$p$ 0.9 TeV   & $0.043\pm0.006$ & $0.070\pm0.004$ & 0.119 \\
            &                             & $p$-$p$ 2.76 TeV  & $0.046\pm0.003$ & $0.064\pm0.002$ & 0.008 \\
            &                             & $p$-$p$ 7 TeV     & $0.053\pm0.007$ & $0.056\pm0.004$ & 0.247 \\
Figure 11(b)&                             & 0-10\%            & $0.019\pm0.004$ & $0.097\pm0.002$ & 0.008 \\
            &                             & 20-30\%           & $0.043\pm0.020$ & $0.075\pm0.009$ & 0.154 \\
            &                             & 40-60\%           & $0.048\pm0.021$ & $0.066\pm0.010$ & 0.187 \\
Figure 11(c)&                             & 0-12\%            & $0.002\pm0.038$ & $0.129\pm0.020$ & 1.693 \\
            &                             & 10-20\%           & $0.001\pm0.035$ & $0.127\pm0.018$ & 1.144 \\
            &                             & 20-40\%           & $0.004\pm0.034$ & $0.124\pm0.018$ & 1.075 \\
            &                             & 40-60\%           & $0.003\pm0.033$ & $0.121\pm0.018$ & 1.197 \\
            &                             & 60-80\%           & $0.005\pm0.032$ & $0.118\pm0.018$ & 1.548 \\
Figure 11(d)&                             & 0-5\%             & $-0.007\pm0.029$& $0.162\pm0.015$ & 1.012 \\
            &                             & 50-60\%           & $-0.055\pm0.040$& $0.174\pm0.023$ & 2.994 \\
            &                             & 80-90\%           & $-0.022\pm0.018$& $0.123\pm0.011$ & 1.006 \\
Figure 11(e)&                             & 0-5\%             & $-0.007\pm0.033$& $0.122\pm0.016$ & 0.716 \\
            &                             & 40-60\%           & $0.036\pm0.011$ & $0.070\pm0.006$ & 0.185 \\
            &                             & 80-100\%          & $0.061\pm0.005$ & $0.019\pm0.005$ & 0.227 \\
\hline\hline
\end{tabular}%
\end{center}
}} }

\newpage
\begin{figure}
\hskip-1.0cm \begin{center}
\includegraphics[width=15.0cm]{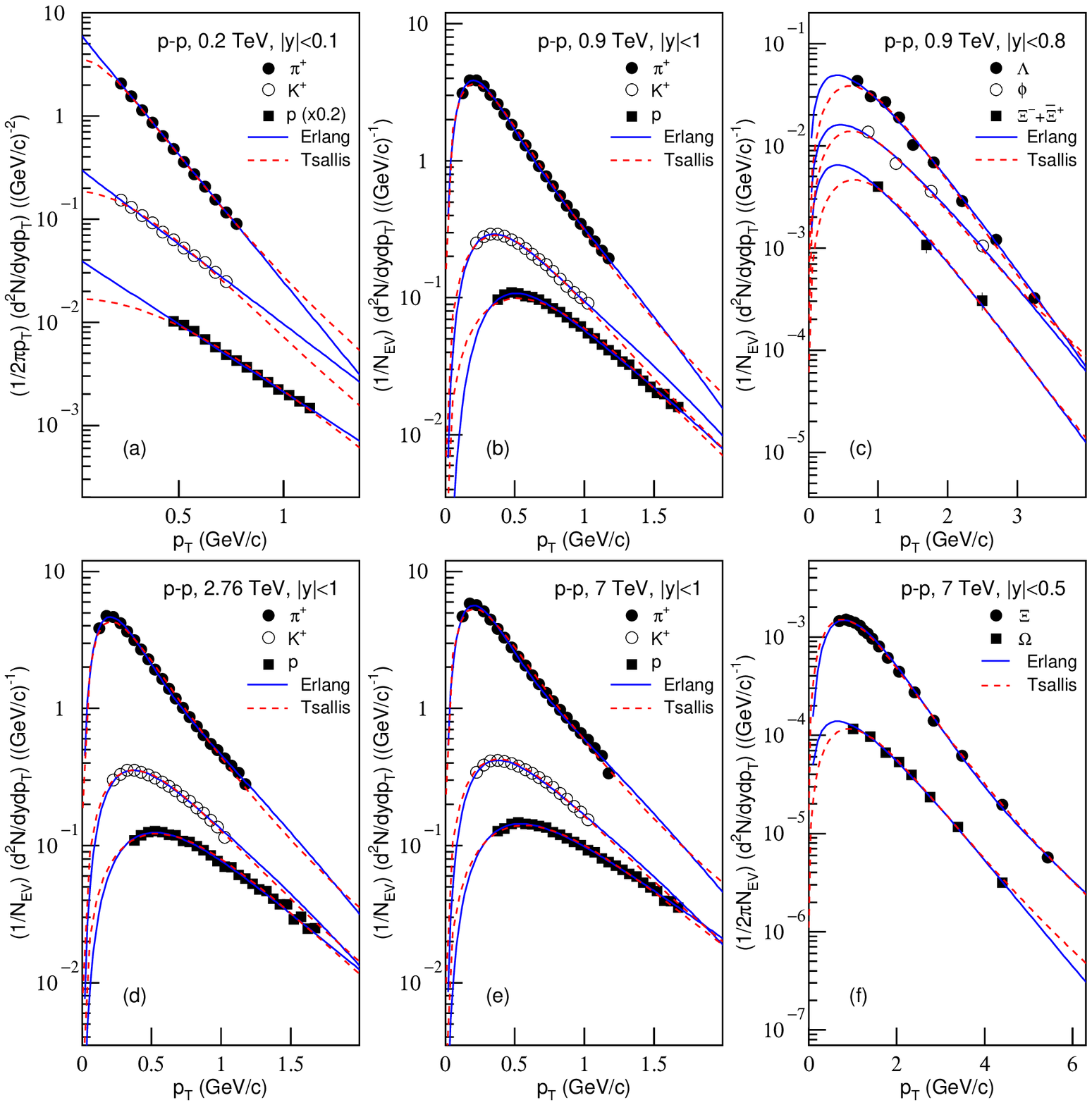}
\end{center}
\vskip1.0cm {\small Figure 1.  Transverse momentum spectra of
various identified hadrons produced in $p$-$p$ collisions at (a)
$\sqrt{s}=0.2$, (b)(c) $0.9$, (d) $2.76$, and (e)(f) $7$ TeV. The
symbols represent the experimental data of (a) $\pi^{+}$, $K^{+}$,
and $p$ measured by the STAR Collaboration at midrapidity
$|y|<0.1$ [15], (b)(d)(e) $\pi^{+}$, $K^{+}$, and $p$ measured by
the CMS Collaboration in the range $|y|<1$ [21], (c) $\Lambda$,
$\phi$, and $\Xi^{-}+\bar{\Xi}^{+}$ measured by the ALICE
Collaboration in the range $|y|<0.8$ [22], as well as (f) $\Xi$
and $\Omega$ measured by the ALICE Collaboration in the range
$|y|<0.5$ [23]. The errors include statistical and systematic
errors. The solid and dashed curves are our results with one- or
two-component Erlang distribution and Tsallis statistics,
respectively. For clarity, the results for $p$ in Figure 1(a) are
scaled by $\times0.2$ shown in the panels.}
\end{figure}

\begin{figure}
\hskip-1.0cm \begin{center}
\includegraphics[width=15.0cm]{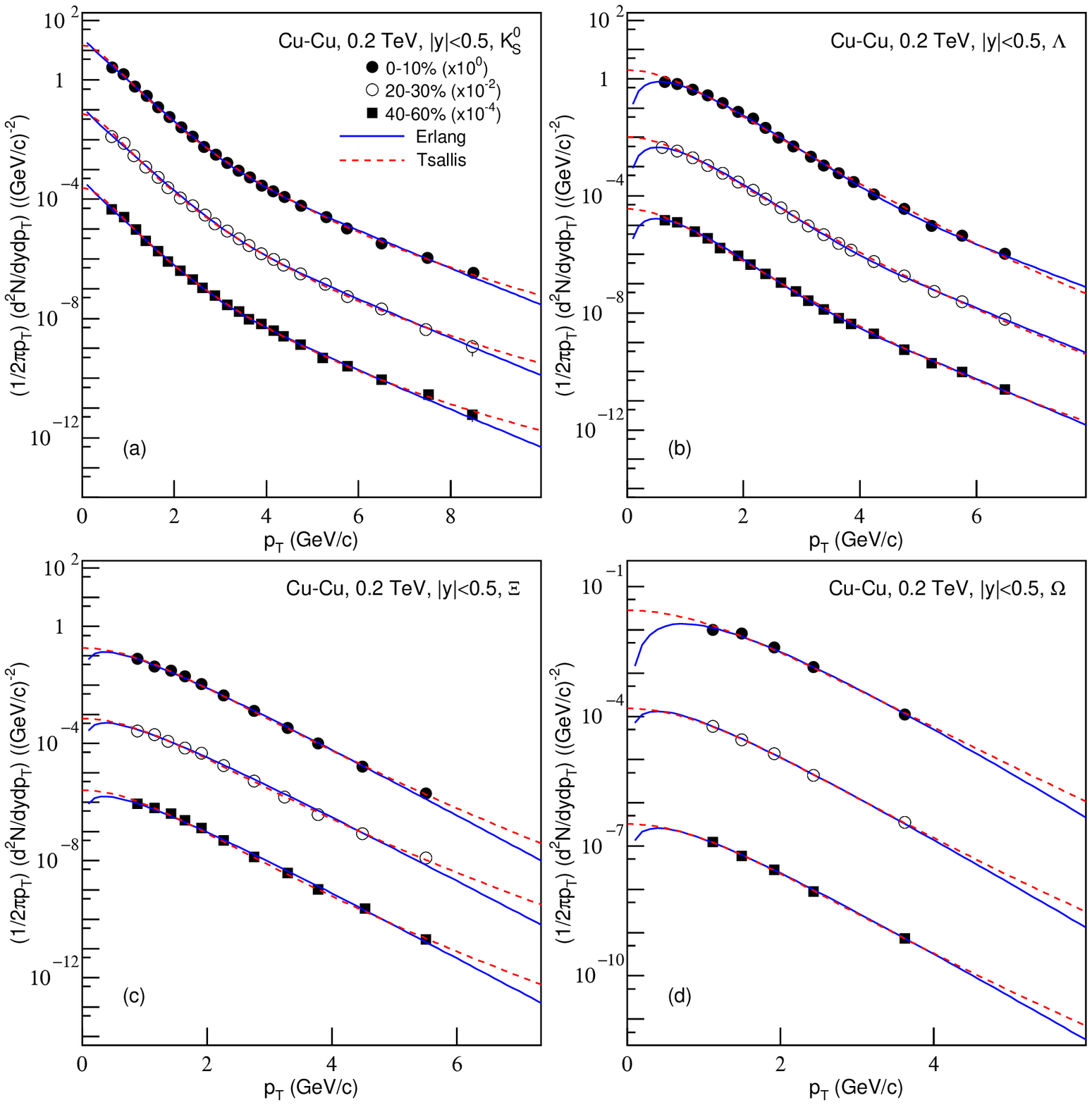}
\end{center}
\vskip1.0cm {\small Figure 2. Transverse momentum spectra of (a)
$K_{s}^{0}$, (b) $\Lambda$, (c) $\Xi$, and (d) $\Omega$ produced
in Cu-Cu collisions at $\sqrt{s_{NN}}=0.2$ TeV. The symbols
represent the experimental data of the STAR Collaboration in
$|y|<0.5$ and different centrality intervals of 0--10\%, 20--30\%,
and 40--60\% [16]. The error bars are combined statistical and
systematic errors. The solid and dashed curves are our results
calculated by using the one- or two-component Erlang distribution
and Tsallis statistics, respectively. For clarity, the results for
different intervals are scaled by different amounts shown in the
panels.}
\end{figure}

\newpage
\begin{figure}
\hskip-1.0cm \begin{center}
\includegraphics[width=15.0cm]{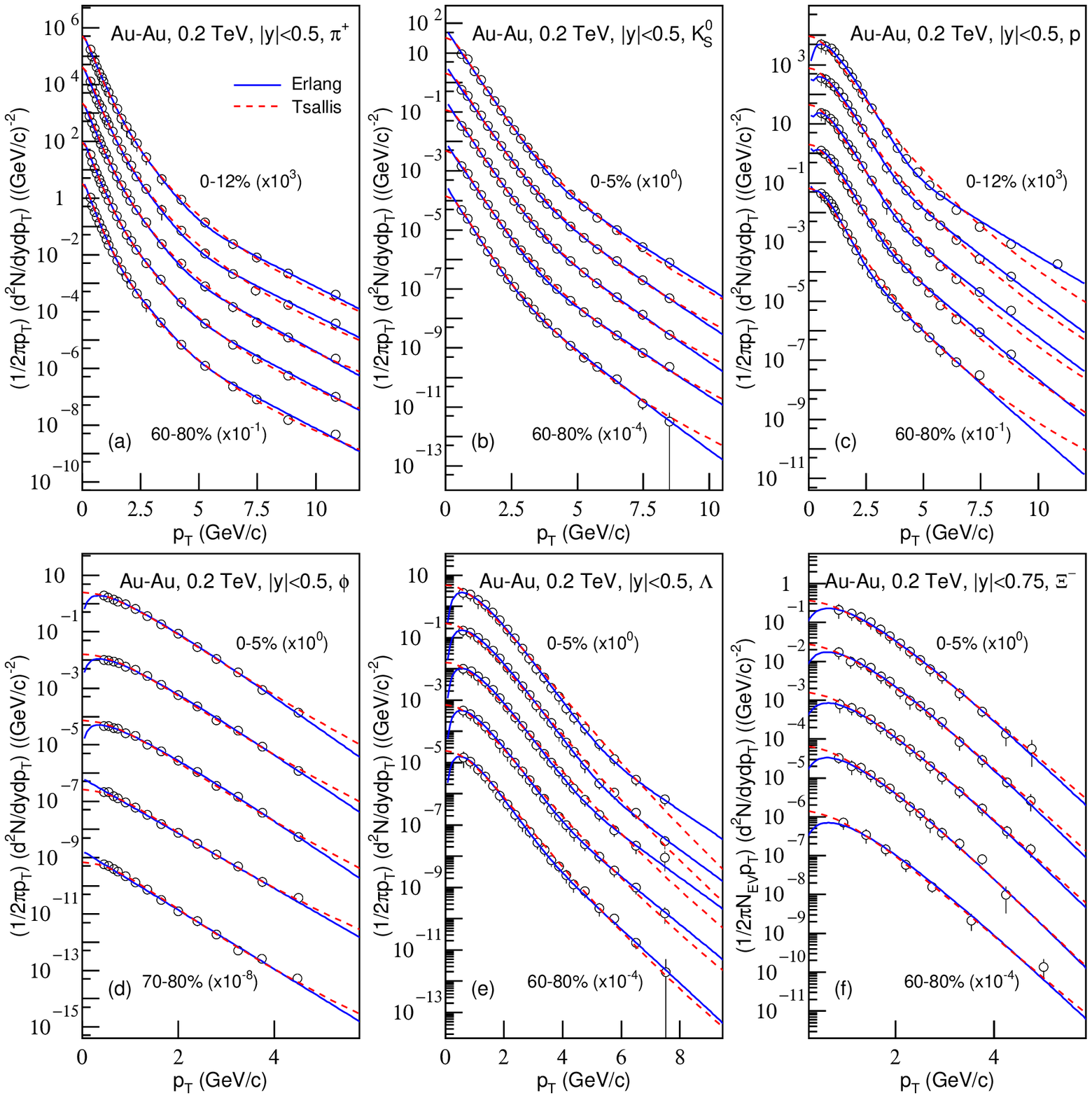}
\end{center}
\vskip1.0cm Figure 3. Transverse momentum spectra of (a) $\pi^+$
for $|y|<0.5$, (b) $K_{S}^{0}$ for $|y|<0.5$, (c) $p$ for
$|y|<0.5$, (d) $\phi$ for $|y|<0.5$, (e) $\Lambda$ for $|y|<0.5$,
and (f) $\Xi^{-}$ for $|y|<0.75$ produced in Au-Au collisions at
$\sqrt{s_{NN}}=0.2$ TeV as a function of centrality. The
experimental data were recorded by the STAR Collaboration [17--20]
and scale factors for different centralities are applied to the
spectra in the panels for clarity. The fitting results with two
types of distributions are plotted by the curves.
\end{figure}

\newpage
\begin{figure}
\hskip-1.0cm \begin{center}
\includegraphics[width=15.0cm]{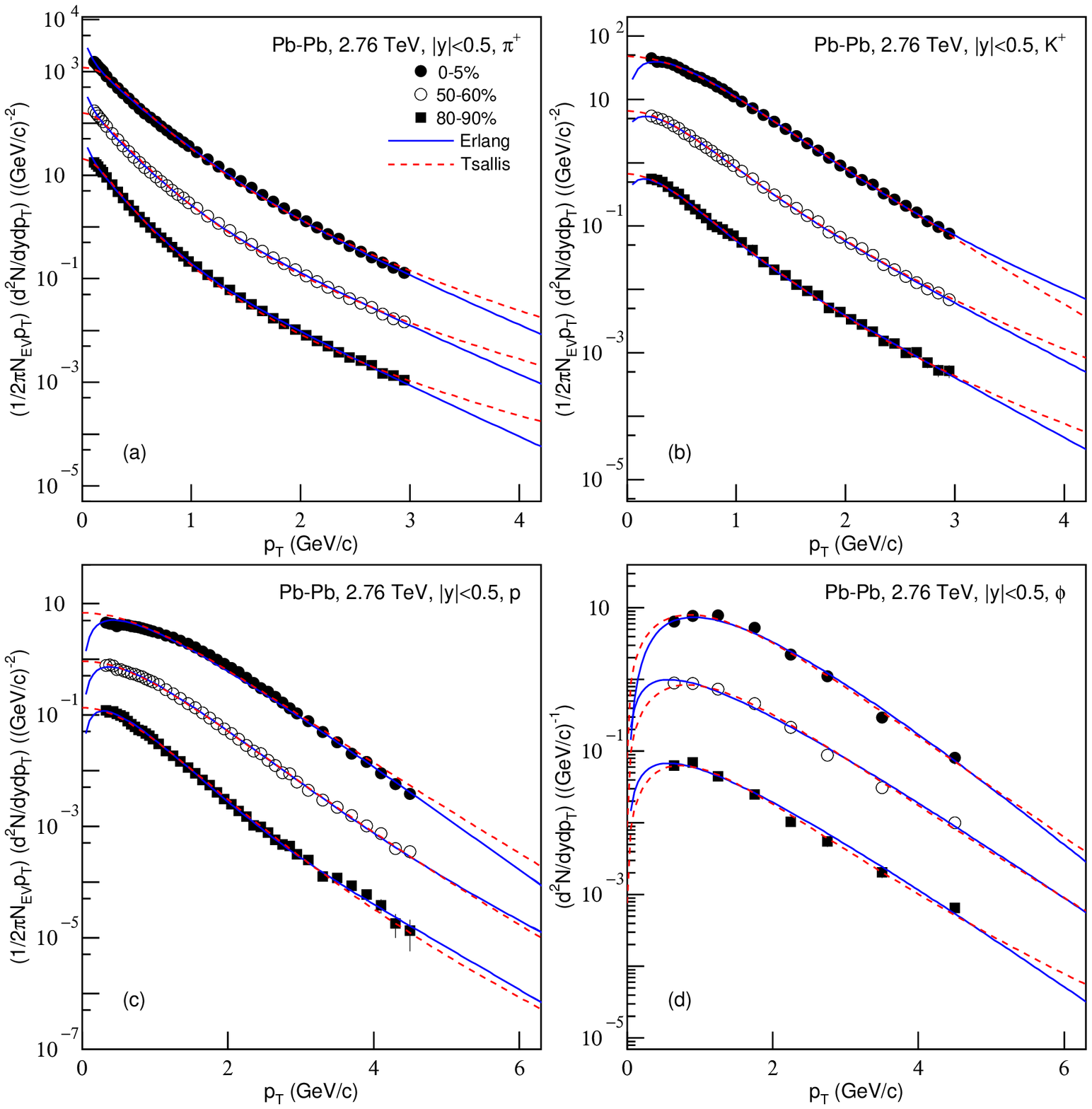}
\end{center}
\vskip1.0cm Figure 4. Transverse momentum spectra of (a)
$\pi^{+}$, (b) $K^{+}$, (c) $p$, and (d) $\phi$ produced in
central (0--5\%), semi-central (50--60\%), and peripheral
(80--90\%) Pb-Pb collisions at $\sqrt{s_{NN}}=2.76$ TeV. The
symbols represent the experimental data measured by the ALICE
Collaboration at midrapidity $|y|<0.5$ [24, 25]. The uncertainties
on the data points are combined statistical and systematic ones.
Our results are exhibited by the curves.
\end{figure}

\newpage
\begin{figure}
\hskip-1.0cm \begin{center}
\includegraphics[width=15.0cm]{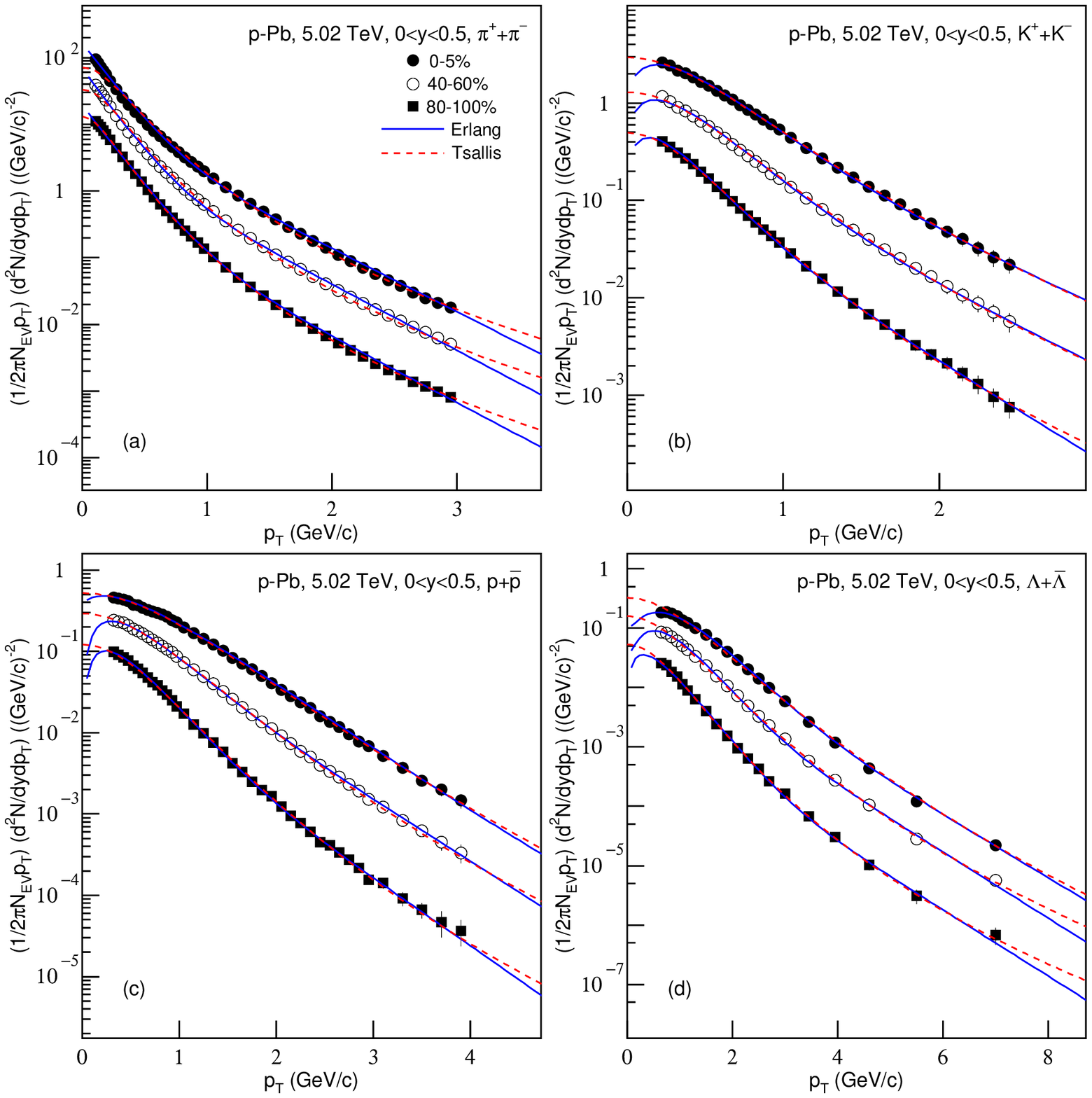}
\end{center}
\vskip1.0cm Figure 5. Transverse momentum spectra of (a) $\pi^{+}
+ \pi^{-}$, (b) $K^{+} + K^{-}$, (c) $p+\bar{p}$, and (d) $\Lambda
+\bar{\Lambda}$ for three centrality bins produced in $p$-Pb
collisions at $\sqrt{s_{NN}}=5.02$ TeV. The ALICE experimental
data in $0<y<0.5$ are represented by different symbols [26]. The
error bars are combined statistical and systematic errors. Our
results analyzed by two-component Erlang distribution and Tsallis
statistics are given by the solid and dashed curves, respectively.
\end{figure}

\newpage
\begin{figure}
\hskip-1.0cm \begin{center}
\includegraphics[width=15.0cm]{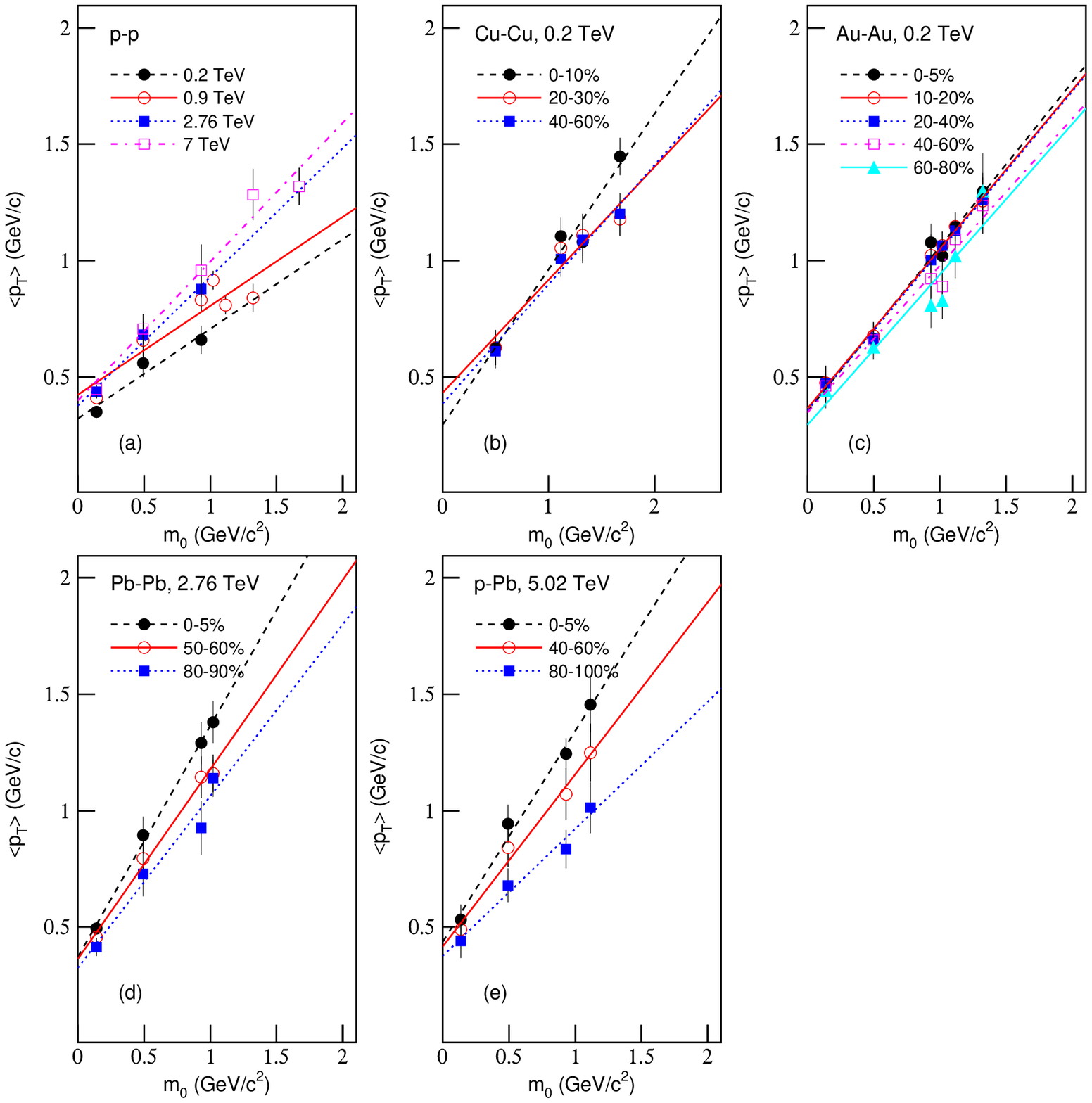}
\end{center}
\vskip1.0cm Figure 6. Rest mass, center-of-mass energy, and
centrality dependences of mean transverse momentum in $p$-$p$,
Cu-Cu, Au-Au, Pb-Pb, and $p$-Pb collisions. The symbols represent
the parameter values extracted from Figures 1-5 and listed in
Tables 1 and 3. The lines represent linear fits of the results as
a function of rest mass using Eq. (7).
\end{figure}

\newpage
\begin{figure}
\hskip-1.0cm \begin{center}
\includegraphics[width=15.0cm]{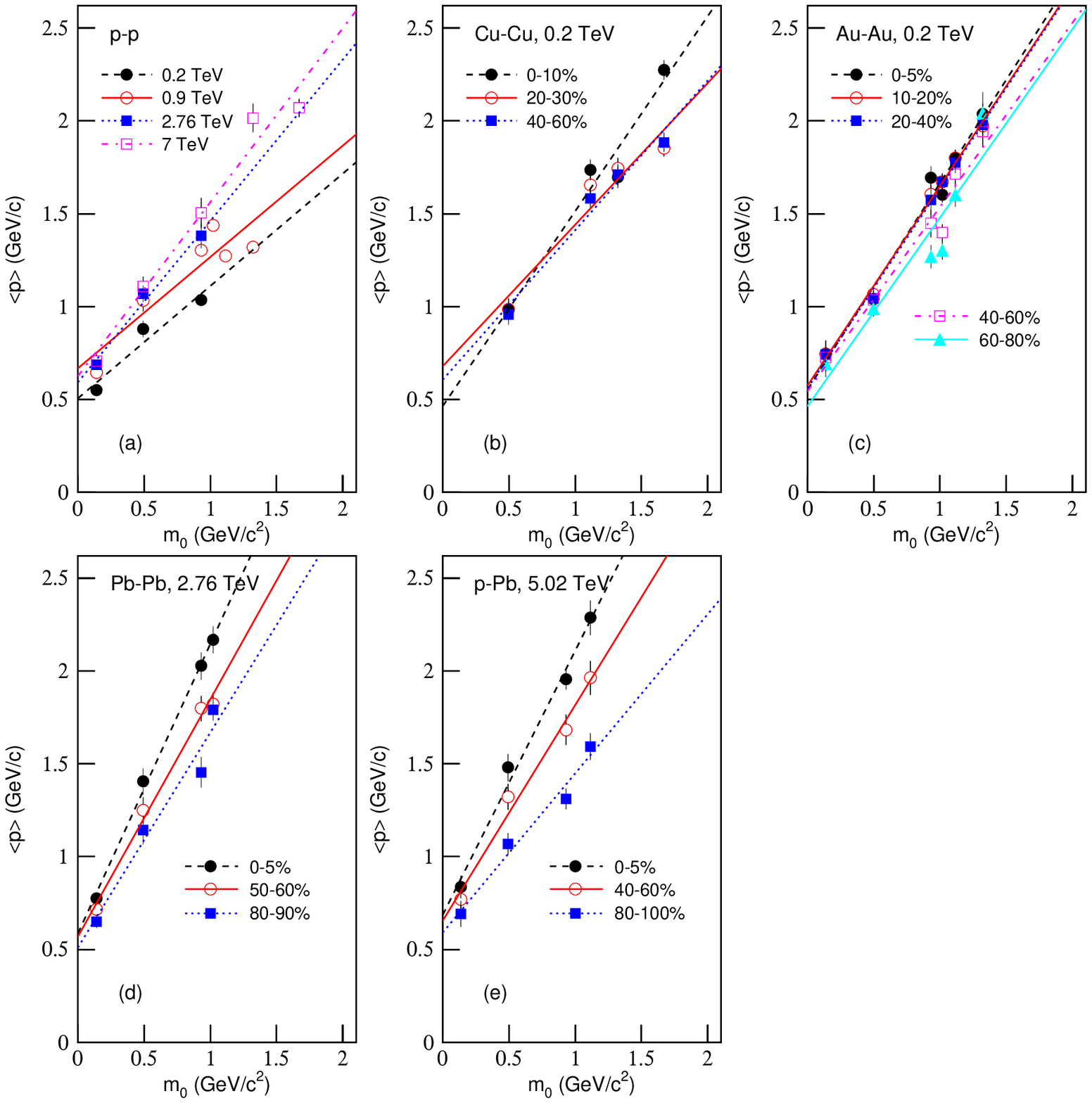}
\end{center}
\vskip1.0cm Figure 7. Rest mass, center-of-mass energy, and
centrality dependences of mean momentum in $p$-$p$, Cu-Cu, Au-Au,
Pb-Pb, and $p$-Pb collisions. The symbols represent the parameter
values extracted from Figures 1--5. The fitting lines are obtained
by using Eq. (8).
\end{figure}

\newpage
\begin{figure}
\hskip-1.0cm \begin{center}
\includegraphics[width=15.0cm]{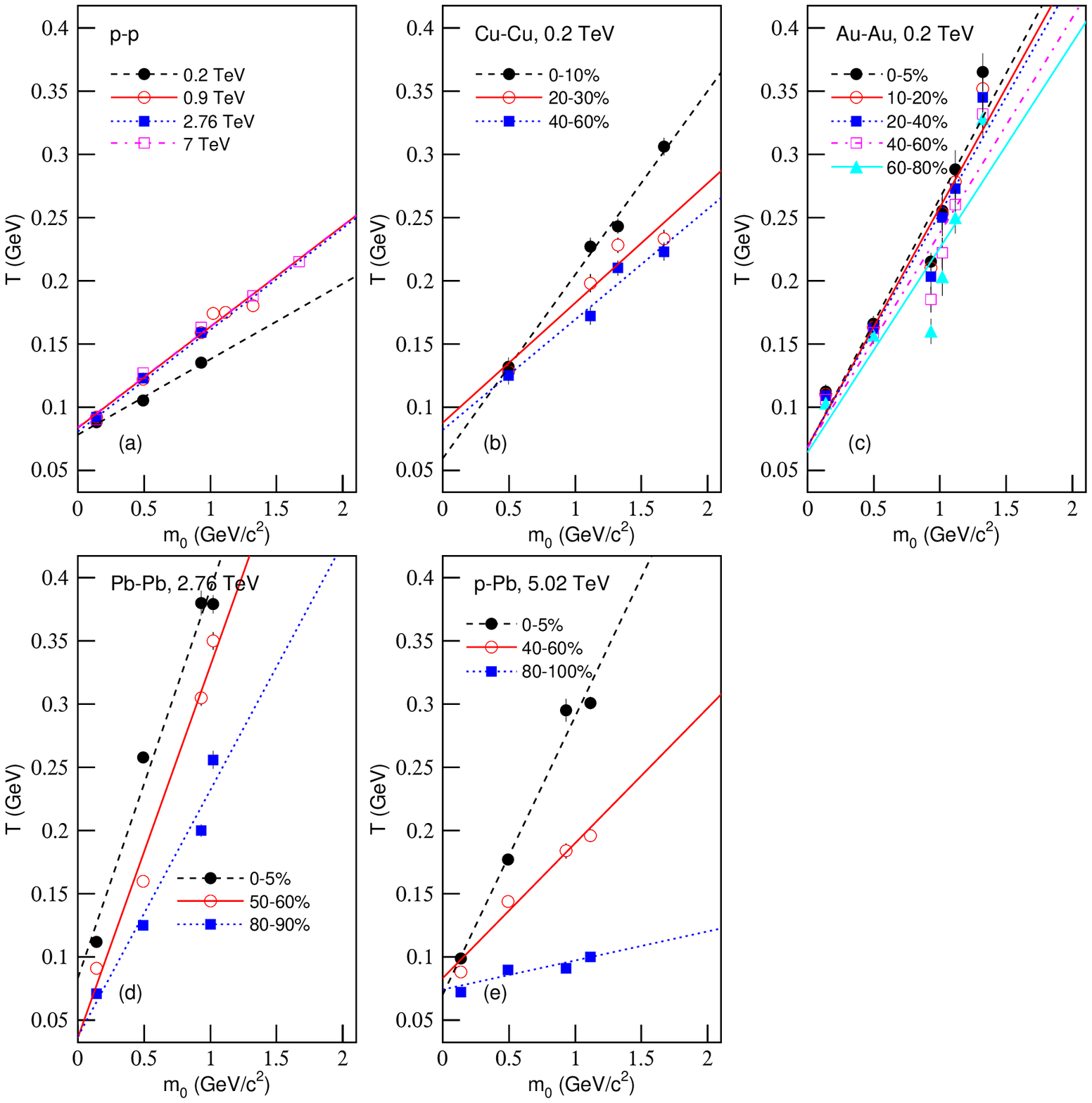}
\end{center}
\vskip1.0cm Figure 8. Dependences of effective temperature on rest
mass, center-of-mass energy, and centrality in $p$-$p$, Cu-Cu,
Au-Au, Pb-Pb, and $p$-Pb collisions.  The symbols represent the
parameter values extracted from Figures 1--5 and listed in Table
2. The fitting lines are obtained by using Eq. (9).
\end{figure}

\newpage
\begin{figure}
\hskip-1.0cm \begin{center}
\includegraphics[width=15.0cm]{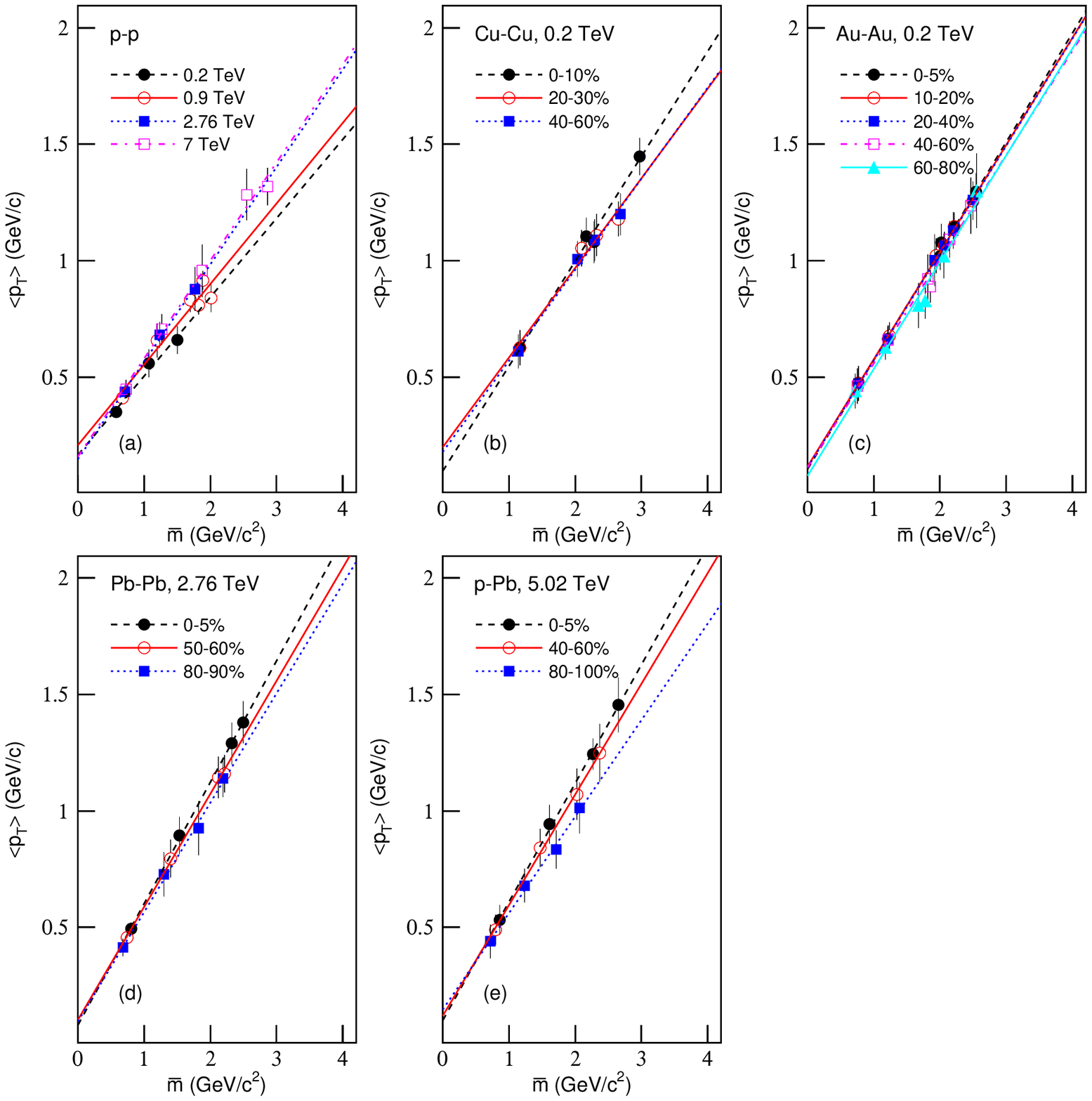}
\end{center}
\vskip1.0cm Figure 9. Moving mass, center-of-mass energy, and
centrality dependences of mean transverse momentum in $p$-$p$,
Cu-Cu, Au-Au, Pb-Pb, and $p$-Pb collisions. The symbols represent
the parameter values extracted from Figures 1-5. The lines
represent linear fits of the results as a function of moving mass
by using Eq. (10).
\end{figure}

\newpage
\begin{figure}
\hskip-1.0cm \begin{center}
\includegraphics[width=15.0cm]{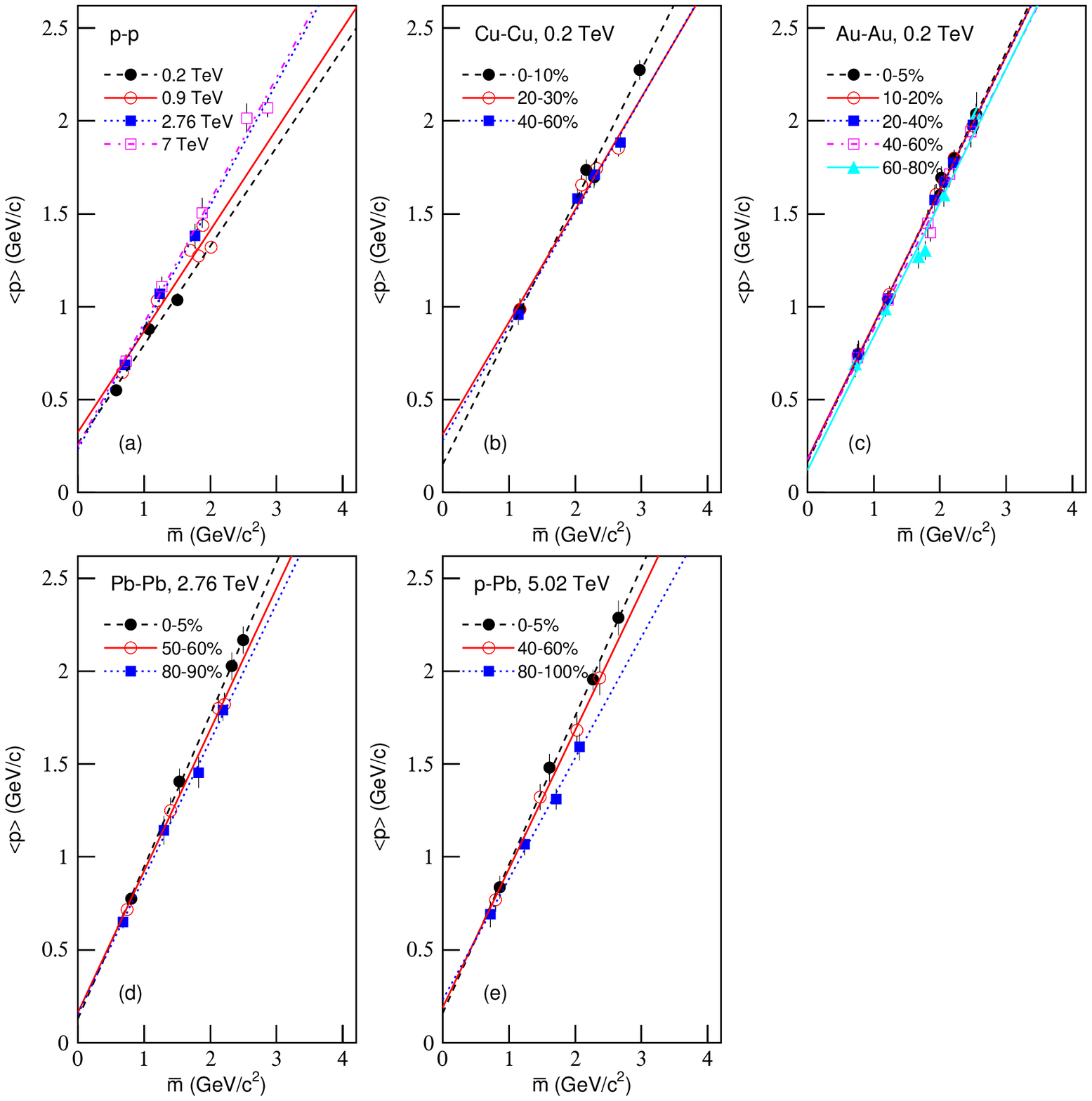}
\end{center}
\vskip1.0cm Figure 10. Moving mass, center-of-mass energy, and
centrality dependences of mean momentum in $p$-$p$, Cu-Cu, Au-Au,
Pb-Pb, and $p$-Pb collisions. The symbols represent the parameter
values extracted from Figures 1--5. The fitting lines are obtained
by using Eq. (11).
\end{figure}

\newpage
\begin{figure}
\hskip-1.0cm \begin{center}
\includegraphics[width=15.0cm]{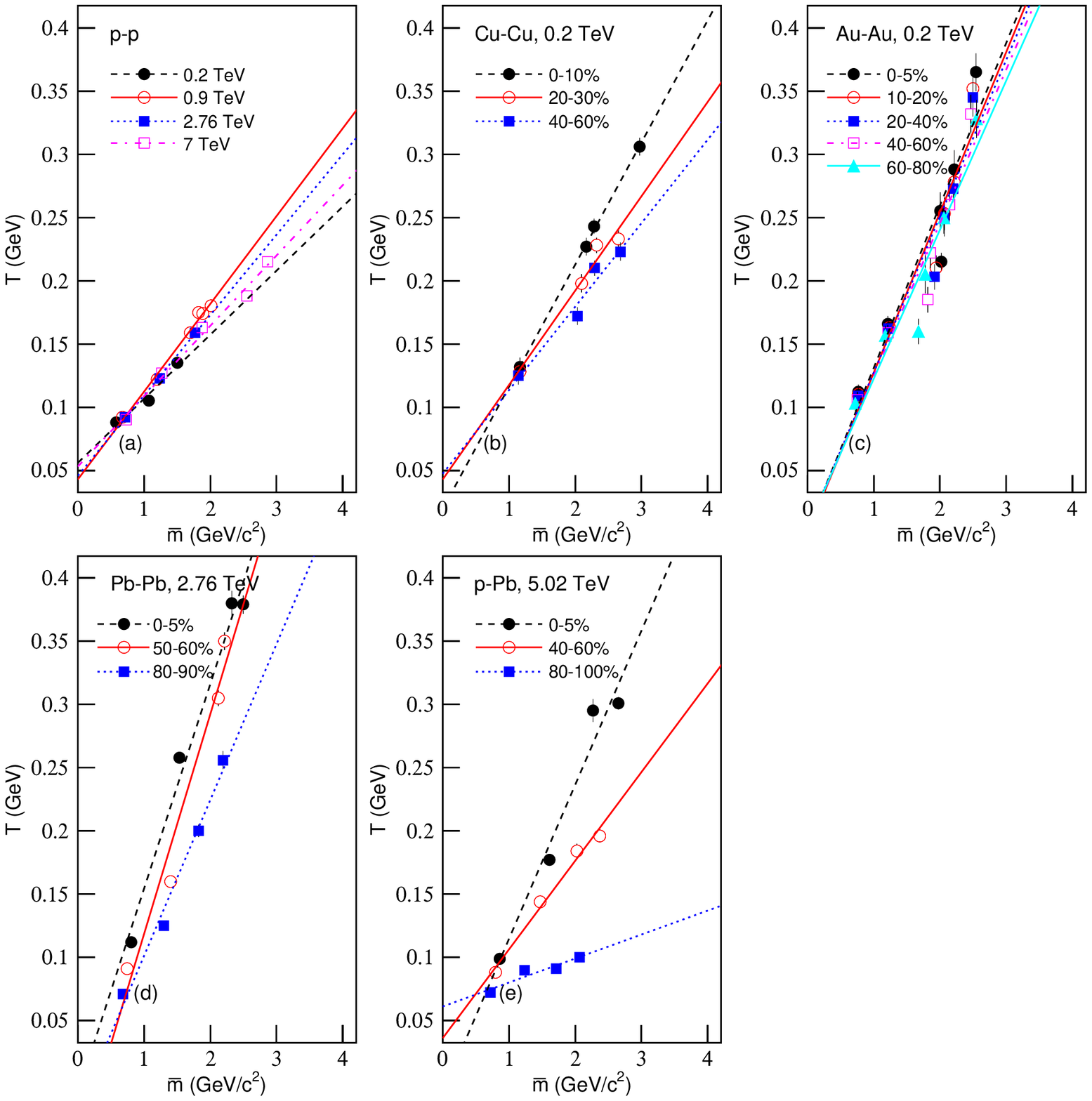}
\end{center}
\vskip1.0cm Figure 11. Dependences of effective temperature on
moving mass, center-of-mass energy, and centrality in $p$-$p$,
Cu-Cu, Au-Au, Pb-Pb, and $p$-Pb collisions. The symbols represent
the parameter values extracted from Figures 1--5 and listed in
Table 2. The fitting lines are obtained by using Eq. (12).
\end{figure}

\newpage
\begin{figure}
\hskip-1.0cm \begin{center}
\includegraphics[width=15.0cm]{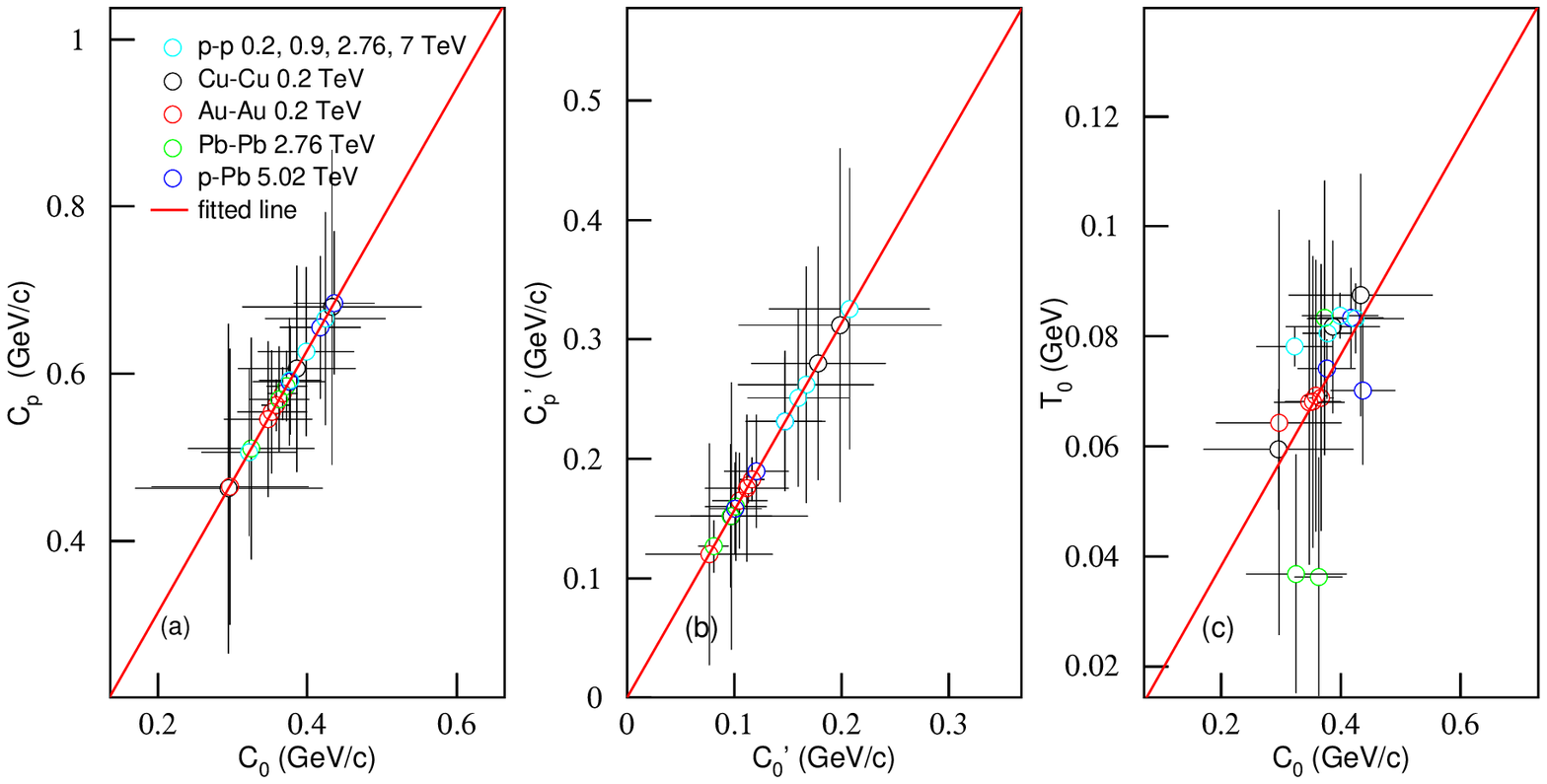}
\end{center}
\vskip1.0cm Figure 12. Correlations between intercept points (a)
$C_{p}-C_{0}$, (b) $C_{p}^{'}-C_{0}^{'}$, and (c) $T_{0}-C_{0}$
from $\langle p_{T}\rangle-m_{0}$, $\langle p \rangle-m_{0}$,
$T-m_{0}$, $\langle p_{T}\rangle-\overline{m}$, and $\langle p
\rangle -\overline{m}$ correlations corresponding to Eqs.
(7)--(11), respectively. The symbols represent the parameter
values extracted from Figures 6--11 and listed in Table 4. The
lines are our fitting results.
\end{figure}

\newpage
\begin{figure}
\hskip-1.0cm \begin{center}
\includegraphics[width=15.0cm]{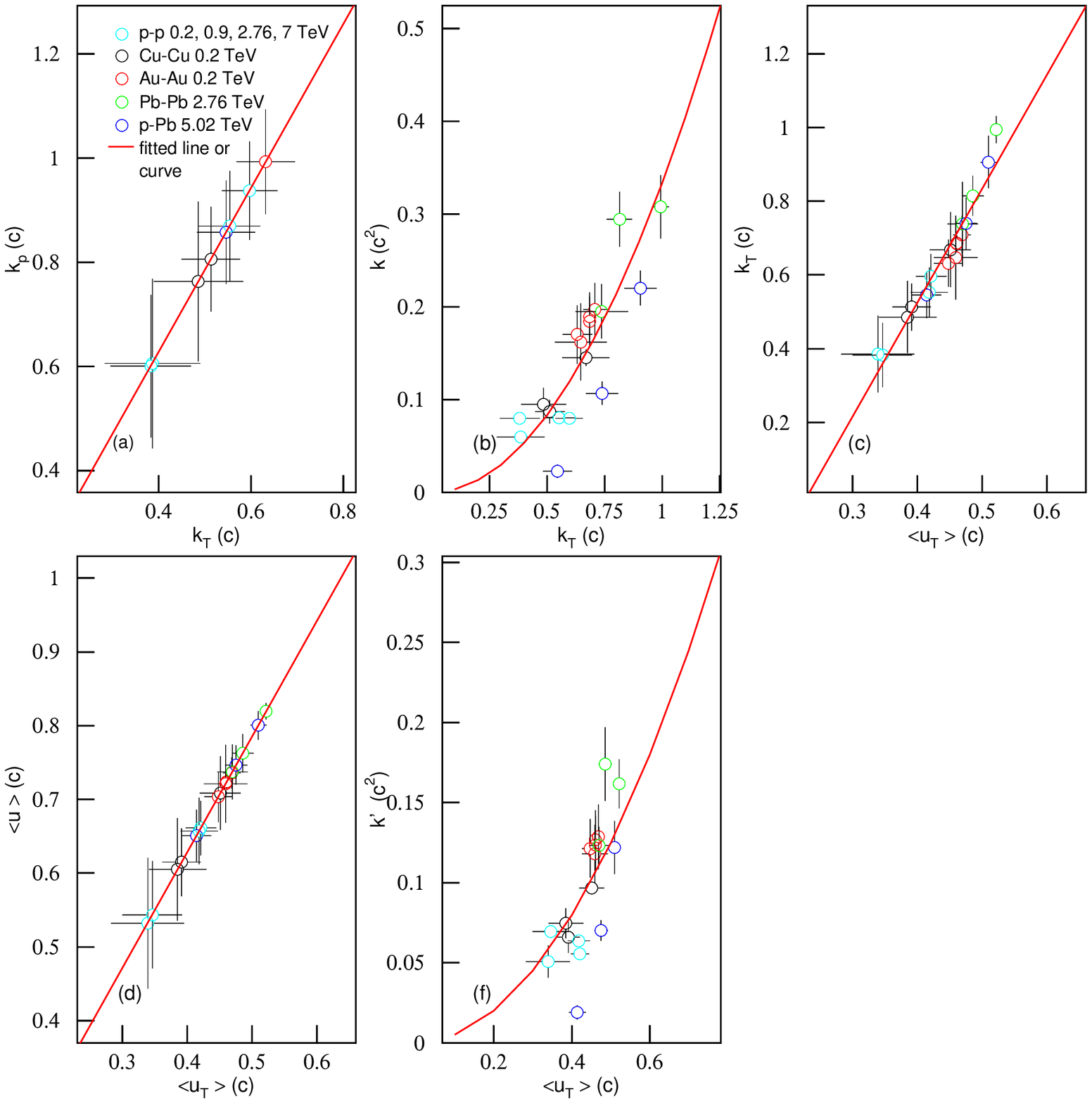}
\end{center}
\vskip1.0cm Figure 13. Correlations between slopes (a)
$k_{p}-k_{T}$, (b) $k-k_{T}$, (c) $k_{T}-\langle u_{T}\rangle$,
(d) $\langle u \rangle-\langle u_{T}\rangle$, and (e)
$k^{'}-\langle u_{T}\rangle$ from $\langle p_{T}\rangle-m_{0}$,
$\langle p \rangle-m_{0}$, $T-m_{0}$, $\langle
p_{T}\rangle-\overline{m}$, $\langle p \rangle -\overline{m}$, and
$T-\overline{m}$ correlations corresponding to Eqs. (7)--(12),
respectively. The symbols represent the parameter values extracted
from Figures 6--11 and listed in Table 4. The lines and curves are
our fitting results.
\end{figure}

\end{document}